\documentstyle[preprint,aps,epsfig,floats]{revtex}

\newcommand{\kv}{{\bf k}_{\perp}}
\newcommand{\lv}{{\bf l}_{\perp}}

\newcommand{\bv}{{\bf b}_{\perp}}
\newcommand{\ac}{ {\centerline{\bf Acknowledgments}} }
\tightenlines
\draft
\begin{document}
\date{\today}
\preprint{{\hbox{RUB-TPII-09/00}}}
\title{Analytic coupling and Sudakov effects in exclusive processes:
       pion and $\gamma ^{*}\gamma \to \pi ^{0}$ form factors \\
       }
\author{N. G. Stefanis,${}^{1}$\thanks{Email:
        stefanis@tp2.ruhr-uni-bochum.de}
        W. Schroers,${}^{2}$\thanks{Email:
        wolfram@theorie.physik.uni-wuppertal.de}
        and
        H.-Ch. Kim ${}^{3}$\thanks{Email:
        hchkim@hyowon.pusan.ac.kr}
        }
\address{${}^{1}$ Institut f\"ur Theoretische Physik II,        \\
                  Ruhr-Universit\"at Bochum,                    \\
                  D-44780 Bochum, Germany                       \\
                  [0.3cm]
         ${}^{2}$ Fachbereich Physik,                           \\
                  Universit\"at Wuppertal,                      \\
                  D-42097 Wuppertal, Germany                    \\
                  [0.3cm]
         ${}^{3}$ Department of Physics,                        \\
                  Pusan National University,                    \\
                  Pusan 609-735, Republic of Korea              \\
         }
\maketitle

\newpage

\begin{abstract}
We develop and discuss in technical detail an infrared-finite
factorization and optimized renormalization scheme for calculating
exclusive processes, which enables the inclusion of transverse
degrees of freedom without entailing suppression of calculated
observables, like form factors. This is achieved by employing an
analytic, i.e., infrared stable, running strong coupling
$\alpha _{\rm s}(Q^{2})$ which removes the Landau singularity at
$Q^{2}=\Lambda _{\rm QCD}^{2}$ by a minimum power-behaved correction.
The ensuing contributions to the cusp anomalous dimension --
related to the Sudakov form factor -- and to the quark anomalous
dimension -- which controls evolution -- lead to an enhancement at high
$Q^{2}$ of the hard part of exclusive amplitudes, calculated in
perturbative QCD, while simultaneously improving its scaling behavior.
The phenomenological implications of this framework are analyzed by
applying it to the pion's electromagnetic form factor, including
the NLO contribution to the hard-scattering amplitude, and also
to the pion-photon transition at LO. For the pion wave function,
an improved ansatz of the Brodsky-Huang-Lepage type is employed,
which includes an effective (constituent-like) quark mass, $m_{\rm
q}=0.33$~GeV. Predictions for both form factors are presented and
compared to the experimental data, applying Brodsky-Lepage-Mackenzie
commensurate scale setting. We find that the perturbative hart part
prevails at momentum transfers above about 20 GeV~${}^{2}$, while at
lower $Q^{2}$-values the pion form factor is dominated by Feynman-type
contributions. The theoretical prediction for the
$\gamma ^{*}\gamma \to \pi ^{0}$ form factor indicates that the true
pion distribution amplitude may be somewhat broader than the asymptotic
one.

\end{abstract}
%Keywords: Renormalization group evolution
%          Perturbative calculations
%          Summation of perturbation theory
%          Sudakov effects
%          Renormalons
%          BLM method of commensurate scale setting
%          Hadron wave functions
%          Electromagnetic meson decays
%          Electromagnetic form factors
\newpage        %finishes title-page
\input amssym.def
\input amssym.tex
\section{I\lowercase{ntroduction}}
\label{sec:intro}

The theoretical description of the QCD running coupling
$\alpha _{\rm s}\left(Q^{2}\right)$
in the low-momentum region has attracted much interest in the last few
years
\cite{Zak92,DMW96,Dok98,SS97,Gru97,Web99,AZ97a,KP95}.
In particular, the possibility of including power corrections into
$\alpha _{\rm s}\left(Q^{2}\right)$,
while preserving its renormalization-group (RG) invariance, enables the
removal of the ghost (Landau) singularity and restores its
$Q^{2}$-analyticity.
Such power corrections, sub-leading in the ultraviolet (UV) region,
correspond to non-analytical contributions to the $\beta$-function as
to make the running coupling well-defined in the infrared (IR) regime
but, being not confined within the UV regime, they are outside the
operator product expansion.

The existence of power corrections in
$\alpha _{\rm s}\left(Q^{2}\right)$,
if true, would greatly affect our understanding of non-perturbative QCD
effects.
For instance, a power correction to $\alpha _{\rm s}$ gives rise to a
linear term in the inter-quark static potential at short distances
\cite{AZ98}.
On a more speculative level, one may argue \cite{AZ97a} that the source
of such terms are small-size fluctuations in the non-perturbative QCD
vacuum, perhaps related to magnetic monopoles in dual QCD or nonlocal
condensates.
Besides, and in practice, a power-behaved contribution at low scales
can be used to remove the Landau singularity, present in perturbation
theory, supplying in this way an IR stable, i.e., ghost-free running
(effective) coupling that can be extended to the timelike region
\cite{Rad82,DMW96,Dok98,SS97,Gru97,BRS00}.
As a result, re-summed expressions, which typically involve integrations
down to scales $\mu \simeq \Lambda _{\rm QCD}$, are not affected by
the Landau pole and can be safely evaluated.

The aim of the present work is to develop in detail a factorization and
renormalization scheme, which self-consistently incorporates such a
non-perturbative power correction in the running coupling, and then use
it to assess and explore exclusive processes.
We do not, however, propose to involve ourselves in the discussion of
whether or not such power corrections have a fundamental justification
within non-perturbative QCD.
We consider the ambiguity in removing the Landau pole as resembling the
ambiguity in adopting a particular (non-IR-finite) renormalization
scheme in perturbative QCD.
However, this scheme dependence will be minimized by combining our
approach with the Brodsky-Lepage-Mackenzie (BLM) commensurate scale
setting procedure \cite{BLM83}.
Recall in this context that the parameter $\Lambda _{\rm QCD}$ has no
special meaning in parameterizing the position of the Landau pole and
can be traded for an interpolating scale $\mu$, on the basis of
the renormalization-scale freedom (see, for instance, \cite{KP95}).
The justification for such an approach will be supplied a posteriori
by the self-consistent incorporation of higher-order perturbative
corrections and by removing the IR-sensitivity of perturbatively
calculated hadronic observables.

A key ingredient of our approach is that the modified running
coupling will be taken into account not only in the factorized
short-distance part, i.e., through the fixed-order perturbation
expansion, but also in the re-summed perturbative expression for
the exponentiation of soft and collinear gluons (Sudakov effects)
and in the RG-controlled evolution of the factorized parts.
This means, in particular, that the exponent of the Sudakov suppression
factor will be generalized to include power corrections, which encode
long-distance physics.

To accomplish these objectives, we adopt as a concrete power-corrected
running coupling an analytic model for $\alpha _{\rm s}$, recently
proposed by Shirkov and Solovtsov \cite{SS97}, which yields an IR-finite
running coupling.
This model combines Lehmann analyticity with the renormalization group
to remove the Landau singularity at $Q^{2}=\Lambda _{\rm QCD}^{2}$,
without employing adjustable parameters, just by modifying the
logarithmic behavior of $\alpha _{\rm s}$ by a (non-perturbative)
minimum power correction in the UV regime.

At the present stage of evidence, it would be, however, premature to
exclude other parameterizations for the behavior of the running
coupling in the infrared, and one could introduce further modifications
\cite{AZ97a,Ale98}.
It is nevertheless worth remarking that in a recent work by Geschkenbein
and Ioffe \cite{GI99} on the polarization operator (related to the Adler
function) the same infrared limit for the effective coupling was
obtained as in the Shirkov-Solovtsov approach.
Furthermore, it was shown in \cite{BRS00} that postulating that the
Adler function $D(Q^{2})$ is given by integrating $R^{\rm QCD}(s)$
over the physical region $s>0$ only, one finds a
$\Lambda$-parameterization for the strong running coupling in the
spacelike region which coincides with the pole-free one-loop expression
proposed by Shirkov and Solovtsov \cite{SS97}.
Hence, the assumption of analyticity of the strong coupling in the
complex $Q^{2}$ plane, used by Shirkov and Solovtsov, which at first
sight might seem arbitrary, is supported by the analyticity of a
physical quantity.

Whatever the particular choice of power corrections in the running
coupling, it is clear, without the Landau pole, IR sensitivity of loop
integrations associated with IR renormalons (see
\cite{Zak92,tHo77,Zak98}, and \cite{Ben98} for a recent review) is
entirely absent.
We emphasize, however, that the two approaches, though they both entail
power-like corrections
$\propto \left(\Lambda _{\rm QCD}/Q\right)^{\rm p}$,
are logically uncorrelated, as the removal of the Landau pole is a
strong-coupling problem (tantamount to defining a {\it universal}
running coupling in the IR region), whereas IR renormalons parameterize
in a {\it process-dependent} way the low-momentum sensitivity in the
re-summation of large-order contributions of the perturbative series of
bubble chains.
In some sense the two approaches appear to have complementary scopes:
employing ``forced analyticity'' of the running coupling attempts to
incorporate non-perturbative input in terms of a power-correction term in
perturbatively calculated entities, like the hard-scattering amplitude
and the Sudakov suppression factor, that are in turn related to
observables (e.g., form factors).
The renormalon technique, on the other hand, tries to deduce as much as
possible about power corrections from (re-summing) perturbation theory.
Whether power corrections inferred from renormalons can link different
processes (universality assumption) is an important question currently
under investigation \cite{DMW96,KS95,AZ95,BB95,NS95}.

Continuing our previous exploratory study \cite{SSK98} (see also
\cite{SSK98a} from which the present investigation partly derives), we
further extend and test our theoretical framework with build-in
analyticity by including into the calculation of the pion form factor
the next-to-leading order (NLO) perturbative contribution to the
hard-scattering amplitude \cite{FGOC81,DR81,BT87,MNP98}.
To compute the pion form factors in the region
$b\sim 1/\Lambda _{\rm QCD}$, where the hadronic size becomes important,
the effects of the original ${\kv}$ distribution of the partons inside
the pion have to be included.
To this end, an ansatz of the Brodsky-Huang-Lepage (BHL) type
\cite{BHL83} for the (soft) pion wave function is adopted.
This ansatz incorporates an effective (i.e., dynamically generated)
quark mass to ensure that $\psi _{\pi}^{\rm soft}\left(x,{\kv}\right)$
has the correct behavior for ${\kv}=0$ and $k_{3}\to -\infty$.
Predictions for both the pion electromagnetic and the pion-gamma
transition form factor, employing a pion light-cone wave function
without a mass term, were presented in Refs.~\cite{SSK98,SSK98a} to
which we refer for further details.
The influence of the mass term on the hard pion form factor is very weak
and it primarily affects the soft contribution and the pion-photon
transition form factor.
Moreover, the specific form of the ansatz for modeling the intrinsic
transverse parton momentum in the pion bound state is insignificant
for the implementation of the IR-finite scheme, as its effects become
relevant only for large transverse distances below
$1/\Lambda _{\rm QCD}$ that are outside the scope of the present
investigation.

To minimize the dependence on the renormalization scheme and scale, we
obtain our results using an optimized renormalization prescription,
based on the Brodsky-Lepage-Mackenzie (BLM) \cite{BLM83} commensurate
scale setting.
The effect of using a commensurate renormalization scale in
calculating $F_{\pi}$ and $F_{\pi\gamma}$ is discussed quantitatively.
An important observation is that the theoretical prediction for the hard
(perturbative) contribution to the pion's electromagnetic form factor
exhibits no IR-sensitivity, in contrast to approaches
\cite{LS92,JK93,TL98,MNP98} with no IR-fixed point in the running
coupling.
The existence of an IR-fixed point in the effective strong coupling is
implied by the general success of the dimensional counting rules.

The major virtue of such a theoretical framework, the latter being the
object of this paper, is that it enables the inclusion of transverse
degrees of freedom, primordial (i.e., intrinsic) \cite{JK93} and those
originating from (soft) gluonic radiative corrections \cite{LS92,BS89},
without entailing suppression of perturbatively calculated observables,
viz., the pion form factor, in the high-momentum region, where the use
of perturbative QCD is justified, and where suppression is merely the
result of an unnecessarily severe IR-regularization.
This enhancement is (as noted above) due to power-term generated
contributions to the anomalous dimensions of the cusped Wilson line,
related to the Sudakov form factor, and such to the quark wave function
that governs evolution of the factorized exclusive amplitude.
These modified anomalous dimensions will be treated here to two-loop
accuracy.
Note that at the same time, the artificial rising trend (see, e.g.,
\cite{MNP98,TL98}) of the magnitude of the hard pion form factor at
intermediate and low momenta (below about 4 GeV${}^{2}$), which solely
originates from the presence of the Landau singularity at
$Q^{2}=\Lambda _{\rm QCD}^{2}$ in the effective coupling, is here
entirely absent.
Therefore, the scaling behavior of the perturbatively calculated pion
form factor towards lower values of $Q^{2}$ resembles the one computed
with a quasi constant coupling.
Indeed, the scaled pion form factor is in a wide range of momenta almost
a straight line, as one should expect for the leading-twist
contribution (modulo logarithmic evolutional corrections which for the
asymptotic distribution amplitude start at NLO and are negligible).
Consequently, the magnitude of the hard part at low $Q^{2}$ is
considerably reduced and falls short to account for the data.
In this momentum regime the pion form factor is dominated by its soft
Feynman-type contribution.

Although most of our considerations refer to the pion as a case study
for the proposed IR-finite framework, the reasoning can be extended to
describe three-quark systems as well.
This will be reported elsewhere.

The outline of the paper is as follows.
In the next section we briefly discuss the essential features of the
IR-finite running coupling.
In Sect.~\ref{sec:IRF} we develop and present in detail our theoretical
scheme.
Sect.~\ref{sec:piffNLO} extends the method to the NLO contribution
to the hard-scattering amplitude.
An important ingredient in the phenomenological analysis of the form
factors is the BHL-type ansatz for modeling
$\psi _{\pi}^{\rm soft}\left(x,{\kv}\right)$,
which includes an effective quark mass.
In Sect.~\ref{sec:valid} we discuss the numerical implementation
of our scheme revolving around the appropriate
kinematic cuts to ensure factorization of dynamical regimes on the
numerical level by appropriately defining the accessible phase space
regions of transverse momenta (or equivalently transverse distances)
for gluon emission in each regime.
In Sect.~\ref{sec:pheno} we apply these techniques to the
electromagnetic and the pion-gamma transition form factor.
We also provide arguments for the appropriate choice of the
renormalization scale and link our renormalization prescription to
BLM optimal, i.e., commensurate, scale setting.
We also discuss how our scheme compares with others.
Finally, in Sect.~\ref{sec:sum}, we summarize our results and draw our
conclusions.

\section{M\lowercase{odel for} QCD \lowercase{running coupling}}
\label{sec:alpha_s}

The key element of the analytic approach of Shirkov and Solovtsov
\cite{SS97} is that it combines a dispersion-relation approach, based
on local duality, with the renormalization group to bridge the regions
of small and large momenta, providing universality at low scales.
The approach is an extension to QCD of a method originally formulated
by Redmond for QED \cite{Red58}.

At the one-loop level, the Landau ghost singularity is removed by a
single power correction and the IR-finite running coupling reads
\begin{eqnarray}
    \bar{\alpha}_{\rm s}^{{\rm an}(1)}(Q^{2})
& \equiv &
    \bar{\alpha}_{\rm s}^{{\rm pert}(1)}(Q^{2})
  + \bar{\alpha}_{\rm s}^{{\rm npert}(1)}(Q^{2})
\nonumber \\
& = &
  \frac{4\pi}{\beta _{0}}
  \left[
          \frac{1}{\ln \left( Q^{2}/\Lambda ^{2} \right)}
        + \frac{\Lambda ^{2}}{\Lambda ^{2} - Q^{2}}
  \right] \; ,
\label{eq:oneloopalpha_an}
\end{eqnarray}
%Eq (1) Analytic model for running QCD coupling constant; 1-loop level
where $\Lambda \equiv \Lambda _{\rm QCD}$ is the QCD scale parameter.

This model has the following interesting properties.
It provides a non-perturbative regularization at low scales and leads
to a universal value of the coupling constant at zero momentum
$
   \bar{\alpha}_{\rm s}^{(1)}(Q^{2}=0)
 = 4 \pi /\beta _{0}
\simeq
   1.396
$
(for three flavors), defined only by group constants.
No adjustable parameters are involved and no implicit ``freezing'',
i.e., no (color) saturation hypothesis of the coupling constant in the
infrared is invoked.

Note that the IR-fixed point (i) does not depend on the scale parameter
$\Lambda$ -- this being a consequence of RG invariance -- and (ii)
extends to the two-loop order, i.e.,
$
 \bar {\alpha}_{\rm s}^{(2)}(Q^{2}=0)
=
 \bar {\alpha}_{\rm s}^{(1)}(Q^{2}=0)
\equiv
 \bar {\alpha}_{\rm s}(Q^{2}=0)
$.
(In the following the bar is dropped.)
Hence, in contrast to standard perturbation theory in a minimal
subtraction scheme, the IR limit of the coupling constant is stable,
i.e., does not depend on higher-order corrections and is therefore
universal.
As a result, the running coupling also shows IR stability.
This is tightly connected to the non-perturbative contribution
$\propto \exp (-4\pi /\alpha \beta _{0})$,
which ensures analytic behavior in the IR domain by eliminating the
ghost pole at $Q^{2} = \Lambda ^{2}$, and extends to higher loop orders.
Besides, the stability in the UV domain is not changed relative to the
conventional approach and therefore UV perturbation theory is preserved.

At very low-momentum values, say, below 1~GeV, $\Lambda _{\rm QCD}$
in this model deviates from that used in minimal subtraction schemes.
However, since we are primarily interested in a region of momenta which
is much larger than this scale, the role of this renormalization-scheme
dependence is only marginal.
In our investigation we use
$
 \Lambda _{\rm QCD}^{{\rm an} (\rm {n_{\rm f}} = 3)}
=
 242~{\rm MeV}
$
which corresponds to
$
 \Lambda _{\rm QCD}^{{\overline{\rm MS}} (\rm {n_{\rm f}} = 3)}
=
 200~{\rm MeV}
$.

The extension of the model to two-loop level is possible, though
the corresponding expression is too complicated to be given explicitly
\cite{SS97}.
An approximated formula -- used in our analysis -- with an inaccuracy
less than $0.5\%$ in the region $2.5 \, \Lambda < Q < 3.5 \, \Lambda$,
and practically coinciding with the exact result for larger values of
momenta, is provided by \cite{SS97}
\begin{equation}
  \alpha _{\rm s}^{{\rm an}(2)}(Q^{2})
=
  \frac{4\pi}{\beta _{0}}
  \left[
          \frac{1}{\ln \frac{Q^{2}}{\Lambda ^{2}}
        + \frac{\beta _{1}}{\beta _{0}^{2}} \,
          \ln \left(
                    1 + \frac{\beta _{0}^{2}}{\beta _{1}}
                        \ln \frac{Q^{2}}{\Lambda ^{2}}
              \right)}
        + \frac{1}{2}\, \frac{1}{1-\frac{Q^{2}}{\Lambda ^{2}}}
        - \frac{\Lambda ^{2}}{Q^{2}}D_{1}
  \right] \; ,
\label{eq:twoloopalpha_an}
\end{equation}
%Eq (2) Approximate expression for two-loop running coupling
where
$
 \beta _{0}
=
 11 - \frac{2}{3}n_{\rm f}
=
 9
$,
$
 \beta _{1}
=
 102 - \frac{38}{3}n_{\rm f}
=
 64
$,
and $D_{1} = 0.035$ for $n_{\rm f} = 3$.

With experimental data at relatively low momentum-transfer values for
most exclusive processes, reliable theoretical predictions based on
perturbation theory are difficult to obtain.
Both the unphysical Landau pole of $\alpha _{\rm s}$ and IR instability
of the factorized short-distance part in the so-called end-point
region are affecting such calculations, especially beyond leading
order (LO).
It is precisely for these two reasons that the Shirkov-Solovtsov
analytic approach to the QCD running coupling can be profitably used
for computing amplitudes describing exclusive processes
\cite{LB80,CZ77,ER80}, like hadronic form factors.
The improvements are then:
(i)   First and foremost, the non-perturbatively generated power
      correction modifies the Sudakov form factor
      \cite{BS89,Col89,KR87,Kor89,GKKS97} via the cusp (eikonal)
      anomalous dimension \cite{Pol79}, and changes also the evolution
      behavior of the soft and hard parts through the modified anomalous
      dimension of the quark wave function.
      This additional contribution to the cusp anomalous dimension is
      the source of the observed IR enhancement (at larger $Q^{2}$
      values) of hadronic observables and helps taking into account {\it
      non-perturbative corrections (power terms in $Q$ and the impact
      parameter $b$) in the perturbative domain}, thus improving the
      quality and scaling behavior of the (perturbative) form-factor
      predictions (at low $Q^{2}$).
      We emphasize in this context that the ambiguity of the Landau
      remover is confined within the momentum regime below the
      factorization scale, whereas above that scale the power correction
      is unambiguous.
      Since we are only interested in the computation of the hard
      contribution in the region
      $Q^{2}\gg {\kv}^{2}$, $Q^{2}\gg m_{\rm q}^{2}$,
      this ambiguity is in fact of minor importance.
(ii)  Factorization is ensured without invoking the additional
      assumption of ``freezing'' the coupling strength in the IR regime
      by introducing, for example, an (external, i.e., {\it ad hoc})
      effective gluon mass in order to saturate color forces at large
      distances, alias, low momenta.
(iii) The Sudakov form factor does not have to serve as an IR protector
      against $\alpha _{\rm s}$ singularities.
      Hence, the extra constraint of using the maximum between the
      longitudinal and the transverse scale as argument of
      $\alpha _{\rm s}$, proposed in \cite{LS92} and used in subsequent
      works, becomes superfluous.
(iv)  The factorization and renormalization scheme we propose on that
      basis enables the optimization of the (arbitrary) constants which
      define the factorization and renormalization scales
      \cite{BS89,Col89,CS81,DS84} -- especially in conjunction with the
      BLM commensurate-scale procedure \cite{BLM83}.
      This becomes particularly important when including higher-order
      perturbative corrections (see, Sect.~\ref{sec:piffNLO}).

\section{I\lowercase{nfrared-finite factorization and renormalization
         scheme}}
\label{sec:IRF}

Application of perturbative QCD is based on factorization, i.e., how
a short-distance part can be isolated from the large-distance physics
related to confinement.
But in order that observables calculated with perturbation theory are
reliable, one must deal with basic problems, like the re-summation of
``soft'' logarithms, IR sensitivity, and the factorization and
renormalization scheme dependence of truncated perturbative expansions.

It is one of the purposes of the present work to give a general and
thorough investigation of such questions, as they are intimately
connected to the behavior of the QCD (effective) coupling at low scales.

The object of our study is the electromagnetic pion's form factor in
the space-like region, which can be expressed as the overlap of the
corresponding full light-cone wave functions between the initial
(``in'') and final (``out'') pion states:
\cite{DY69,LBHM83}
\begin{equation}
  F_{\pi}\left(Q^{2}\right)
=
  \sum_{n, \lambda _{i}}^{}
  \sum_{q}^{}
             e_{\rm q}
  \int_{}^{}
             \frac{[dx_{i}][d^{2}{\bf k}_{\perp i}]}{16\pi ^{3}}
  \psi _{\pi}^{\rm out}\left(
                             x_{i}, {\bf l}_{\perp i}, \lambda _{i}
                       \right)
  \psi _{\pi}^{\rm in} \left(
                             x_{i}, {\bf k}_{\perp i}, \lambda _{i}
                       \right)
\label{eq:DYW}
\end{equation}
%Eq (3) Pion form factor as direct overlap of wave functions
%       (DYW formula)
with
\vbox{\begin{mathletters}
\label{measures}
\begin{equation}
  [dx_{i}]
\equiv
  \prod_{i}^{} dx_{i}
  \delta \left( 1 - \sum_{j}^{} x_{j} \right)
\label{eqa}
\end{equation}
\begin{equation}
  [d^{2}{\bf k}_{\perp i}]
\equiv
  \prod_{i}^{} d^{2}{\bf k}_{\perp i} 16 \pi ^{3}
  \delta ^{2}\left( \sum_{j}^{} {\bf k}_{\perp j} \right) \; ,
\label{eqb}
\end{equation}
\end{mathletters}}
%Eq (4) Integration measures
where the sum in Eq.~(\ref{eq:DYW}) extends over all Fock states
and helicities $\lambda _{i}$ (with $e_{\rm q}$ denoting the charge of
the struck quark), and where
\begin{equation}
  {\bf l}_{\perp i}
=
  \left\{\begin{array}{ll}
        {\bf k}_{\perp i} +(1-x_{i}){\bf q}_{\perp i} \; ,
  \;\;\;\;\;\;\;\;\;\; & {\rm struck} \;\, {\rm quark} \\
        {\bf k}_{\perp i} - x_{i}{\bf q}_{\perp i}
  \;\;\;\;\;\;\;\;\;\; & {\rm spectators} \; .
\end{array}
\right.
\label{eq:transmom}
\end{equation}
%Eq (5) Transverse momenta of outgoing quarks
We will evaluate expression (\ref{eq:DYW}) using only the valence
(i.e., lowest particle-number) Fock-state wave function,
$\psi _{\rm q\bar{\rm q}}(x, \kv )$, which provides the leading
twist-2 contribution, since higher light-cone Fock-state wave functions
require the exchange of additional hard gluons and are therefore
relatively suppressed by inverse powers of the momentum transfer
$Q^{2}$.
Furthermore, a recent study \cite{BKM99}, based on light-cone sum rules,
shows that, for the asymptotic pion distribution amplitude, the twist-4
contribution to the scaled pion form factor
($Q^{2}F_{\pi}\left(Q^{2}\right)$
is less than 0.05, whereas the twist-6 correction turns out to be
negligible.
As we shall see in Sect.~\ref{sec:valid}, this higher-twist correction
amounts to about $25\%$ of the NLO hard contribution, calculated in our
scheme.
This uncertainty in the theoretical prediction is much lower than the
quality of the currently available experimental data.

In order to apply a hard-scattering analysis, we dissect the pion wave
function into a soft and a hard part with respect to a factorization
scale $\mu_{\rm F}$, separating the perturbative from the
non-perturbative regime, and write (in the light-cone gauge $A^{+}=0$)
\begin{equation}
    \psi _{\pi}\left( x, \kv \right)
 =
    \psi _{\pi}^{\rm soft}\left( x, \kv \right)
    \theta \left( \mu _{\rm F}^{2} - \kv ^{2} \right)
  + \psi _{\pi}^{\rm hard}\left( x, \kv \right)
    \theta \left( \kv ^{2} - \mu _{\rm F}^{2} \right) \; ,
\label{eq:psidec}
\end{equation}
%Eq (6) Decomposition of pion wave function into soft and hard part
where the wave function $\psi _{\pi} (x, \kv )$ is the amplitude for
finding a parton in the valence Fock state with longitudinal momentum
fraction $x$ and transverse momentum $\kv$ (we suppress henceforth
helicity labels).
Then the large (perturbative) $k_{\perp}$ tail can be extracted from
the soft wave function via a single-gluon exchange kernel, encoded in
the hard scattering amplitude $T_{\rm H}$, so that \cite{LB80,LBHM83}
\begin{equation}
  \psi _{\pi}^{\rm hard}\left( x, \kv \right)
=
  \int_{0}^{1} dy \int_{}^{}d^{2}\lv
  T_{\rm H} \left( x, {\kv}; y, {\lv}\right)
  \psi _{\pi}^{\rm soft}\! \left( y, \lv\right) \; .
\label{eq:hardgluonpot}
\end{equation}
%Eq (7) Extraction of hard tail of wave function via 1-gluon exchange
%       kernel
As a result, the pion form factor in Eq.~(\ref{eq:DYW}) can be
expressed in the factorized form
\begin{eqnarray}
\everymath{\displaystyle}
   F_{\pi} (Q^{2})
= &&
   \psi _{\rm soft}^{\rm out}
   \otimes
   \psi _{\rm soft}^{\rm in}
 + \psi _{\rm soft}^{\rm out}
   \otimes
   \left[ T_{\rm H} \otimes
          \psi _{\rm soft}^{\rm in}
   \right]
 +
   \left[ \psi _{\rm soft}^{\rm out} \otimes
          T_{\rm H}
   \right]
   \otimes
   \psi _{\rm soft}^{\rm in}
\nonumber \\
&& +
     \left[ \psi _{\rm soft}^{\rm out} \otimes
            T_{\rm H}
     \right]
     \otimes
     \left[ T_{\rm H} \otimes
            \psi _{\rm soft}^{\rm in}
     \right] + \ldots \; ,
\label{eq:softhard}
\end{eqnarray}
%Eq (8) Dissection of pion form factor in soft and hard part
where the symbol $\otimes$ denotes convolution defined by
Eq.~(\ref{eq:hardgluonpot}).
The first term in this expansion is the soft contribution to the form
factor (with support in the low-momentum domain) that is not computable
with perturbative methods.
The second term represents the leading-order hard contribution due
to one-gluon exchange, whereas the last one gives the NLO correction,
and the ellipsis represents still higher-order terms.
We will not attempt to derive the first term from non-perturbative QCD,
but we shall adopt for simplicity the phenomenological approach proposed
by Kroll and coworkers in \cite{JKR96} (see also \cite{JK93}),
including, in particular, an effective (constituent-like) quark mass
in the soft pion wave function for the reasons already mentioned in
the introduction and based on arguments to be given below.
This leads to a significantly stronger fall-off with $Q^{2}$ of the
soft contribution to the space-like pion form factor, compared to their
approach, whereas the hard part remains almost unaffected -- as
should be the case if factorization of hadronic size effects is
preserved.
Though Jakob and Kroll consider in \cite{JK93} the option of a Gaussian
$k_{\perp}$-distribution with $m_{\rm q}\neq 0$ and argue that in that
case $F_{\pi}^{\rm soft}$ is significantly reduced, they do not present
predictions for the form factor and do not follow this option any
further (see, however, the predictions in Ref.~\cite{JKR96}).
For other, more sophisticated, attempts to model the soft contribution
to $F_{\pi}(Q^{2})$, we refer to \cite{BKM99,Rad84,Szc98,BRGS00}.

\begin{figure}
\tighten
\[
\psfig{figure=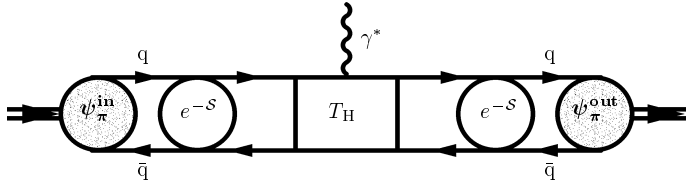,%
       bbllx=136pt,bblly=688pt,bburx=485pt,bbury=753pt,%
       width=10.5cm,silent=}
\]
\vspace{1cm}
\caption[fig:feynmangraph]
        {\tenrm Illustration of the factorized pion form factor,
         exhibiting the different regimes of dynamics.
         The wiggly line denotes the off-shell photon.
         Gluon exchanges are not explicitly displayed.
         The region of hard-gluon re-scattering (LO and NLO) is contained
         in the short-distance part, termed $T_{\rm H}$.
         The blobs ${\rm e}(-S)$ represent in axial gauge Sudakov-type
         contributions, whereas non-perturbative effects are absorbed
         into the (universal) pion wave functions
         $\psi _{\pi}^{\rm in}$ and $\psi _{\pi}^{\rm out}$.
\label{fig:feypi}}
\end{figure}

We now employ a modified factorization prescription \cite{LS92,JK93},
which explicitly retains transverse degrees of freedom, and define
(see for illustration Fig.~\ref{fig:feypi})
\begin{equation}
  \psi _{\pi}^{\rm hard}
=
  \psi _{\pi}^{\rm soft}
  \left(
        \kv ^{2}\leq \frac{C_{3}^{2}}{b^{2}}
  \right)
  \exp \left[ - S \left( \frac{C_{1}^{2}}{b^{2}}
                         \leq \kv ^{2}
                         \leq C_{2}^{2}\xi ^{2}Q^{2}
                  \right)
       \right]
  T_{\rm Hard}\left( Q^{2} \geq \kv ^{2}
                     \geq \frac{C_{3}^{2}}{b^{2}}
              \right) \; ,
\label{eq:modpsi}
\end{equation}
%Eq (9) Modified factorization of pion wave function
with $b$, the variable conjugate to $k_{\perp}$, being the transverse
distance (impact parameter) between the quark and the anti-quark in the
pion valence Fock state.
The Sudakov-type form factor $\exp (-S)$ comprises leading and
next-to-leading logarithmic corrections, arising from soft and
collinear gluons, and re-sums all large logarithms in the region where
$\Lambda _{\rm QCD}^{2}\ll\kv ^{2}\ll Q^{2}$ \cite{CS81,DS84,DDT80}.
The source of these logarithms is due to the incomplete cancellation
between soft-gluon bremsstrahlung and radiative corrections.
It goes without saying that the function $S$ includes
anomalous-dimension contributions to match the change in the running
coupling in a commensurate way with the changes of the renormalization
scale (see below for more details).

%%%%%%%%%%%%%%%%%%%%%%%%%%%%%%%%%%%%%%%%%%%%%%%%%%%%%%%%%%%%%%%%%%%%%
%                            F I G U R E  1                         %
%                          \label{fig:feypi}                        %
%%%%%%%%%%%%%%%%%%%%%%%%%%%%%%%%%%%%%%%%%%%%%%%%%%%%%%%%%%%%%%%%%%%%%
%
%%%%%%%%%%%%%%%%%%%%%%%%%%%%%%%%%%%%%%%%%%%%%%%%%%%%%%%%%%%%%%%%%%%%%

Going over to the transverse (or impact) configuration space (typical
in eikonalization procedures), the pion form factor reads
\cite{LS92}\footnote{Note that this expression cannot be directly
derived from Eq.~(\ref{eq:DYW}).}
\begin{eqnarray}
\everymath{\displaystyle}
  F_{\pi}\left(Q^{2}\right)
= &&
  \int_{0}^{1} dx dy \int_{-\infty}^{\infty}
                     \frac{d^{2}\bv}{(4\pi )^{2}} \,
          {\cal P}_{\pi}^{\rm out}
          \left( y, b, P^{\prime}; C_{1}, C_{2}, C_{4}
          \right)
  T_{\rm H}\left(
                 x, y, b, Q; C_{3}, C_{4}
           \right)
\nonumber \\
&&
\times\;  {\cal P}_{\pi}^{\rm in}
                   \left( x, b, P; C_{1}, C_{2}, C_{4}
                   \right) \; ,
\label{eq:piffbspace}
\end{eqnarray}
%Eq (10) Pion form factor in transverse configuration space
where the modified pion wave function is defined in terms of matrix
elements, viz.,
\begin{eqnarray}
  {\cal P}_{\pi}
                \left( x, b, P, \mu
                \right)
& = &
  \int_{}^{|\kv |<\mu} d^{2}\kv {\rm e}^{- i \kv \cdot \bv}
  {\tilde{\cal P}}_{\pi}\left( x, \kv , P \right)
\nonumber \\
& = &
  \int_{}^{} \frac{dz^{-}}{2\pi}
  {\rm e}^{ -ix P^{+}z^{-}}
  {\left\langle 0 \left\vert
  {\rm T} \left(
                \bar{q}(0)\gamma ^{+}\gamma _{5}
                q\left(0,z^{-},\bv \right)
          \right)
  \right\vert \pi (P) \right\rangle}_{A^{+}=0}
\label{eq:matrel}
\end{eqnarray}
%Eq (11) Definition of modified pion wave function through matrix
%        elements evaluated on the light cone
with $P^{+}=Q/\sqrt{2}=P^{-\prime}$, $Q^{2}=-(P^{\prime}-P)^{2}$,
whereas the dependence on the renormalization scale $\mu$ on the rhs
of Eq.~(\ref{eq:matrel}) enters through the normalization scale of the
current operator evaluated on the light cone and the dependence on the
effective quark mass has not been displayed explicitly.
Note that in the light-cone gauge $A^{+}=0$, the Schwinger string
in Eq.~(\ref{eq:matrel}) reduces to unity.
The factorized hard part
$
 T_{\rm H}\left(
                 x, y, b, Q; C_{3}, C_{4}
          \right)
$
contains hard-scattering quark-gluon subprocesses, including in the
gluon propagators power-suppressed corrections due to their
transverse-momentum dependence.
These gluonic corrections become important in the end-point region
($x\to 0$) for fixed $Q^{2}$.
Furthermore, current quark masses, being much smaller than the
resolution scale (set by the invariant mass of the partons) can be
safely neglected in $T_{\rm H}$, so that (valence) quarks are treated
on-mass shell.\footnote{Furthermore the chiral limit is adopted here,
i.e., $M_{\pi}=0$, since the pion mass is much smaller than the typical
normalization scale in Eq.~(\ref{eq:matrel}).}

\begin{figure}
\tighten
\[
\psfig{figure=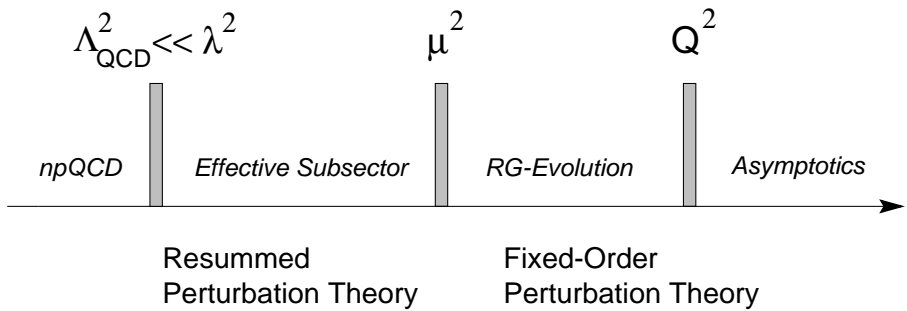,%
       bbllx=0pt,bblly=0pt,bburx=414pt,bbury=142pt,%
       width=12cm,silent=}
\]
\vspace{-0.5cm}
\caption[fig:regimes]
        {\tenrm Regimes of chromodynamics characterized by scales
         typical for the quanta involved.
\label{fig:qcdreg}}
\end{figure}

A few comments on the scales involved and corresponding dynamical
regimes (see Eq.~(\ref{eq:piffbspace}) and Fig.~\ref{fig:qcdreg}) are
in order:
\begin{itemize}
\item The scale $C_{3}/b$ serves to separate perturbative from
      non-perturbative transverse distances (lower factorization scale
      of the effective sub-sector and transverse cutoff).
      We assume that some characteristic scale
      $
       b_{\rm nonp}^{-1} \simeq
       \langle \kv ^{2}\rangle ^{1/2}/x(1-x) \simeq 0.5
      $~GeV
      exists, related to the typical virtuality (off-shellness) of
      vacuum quarks.
      This scale should also provide the natural starting point
      for the evolution of the pion wave function.
      In the following, we match the non-perturbative scale $C_{3}/b$
      with the scale $C_{1}/b$, where the re-summation of soft gluons
      in the effective sub-sector starts, i.e., we set $C_{1}=C_{3}$.
      The lowest boundary of the scale $C_{1}/b$ (IR cutoff) is set by
      $\Lambda _{\rm QCD}$, though the results are not very sensitive
      to using a somewhat larger momentum scale, as we shall see later.
\item The re-summation range in the Sudakov form factor is limited from
      above by the scale $C_{2}\xi Q$ (upper factorization scale of the
      effective sub-sector and collinear cutoff).\footnote{Note that the
      constant $C_{2}$ here differs in notation by a factor of
      $\sqrt{2}$ relative to that used by Collins, Soper, and Sterman in
      \cite{CS81}, i.e.,
      $C_{2}^{\rm CSS} = \sqrt{2}C_{2}$.}
      This scale may be thought of as being an UV-cutoff for the
      effective sub-sector, i.e., for the Sudakov form factor, and
      enables this way a RG-controlled scale dependence governed by
      appropriate anomalous dimensions within this sub-sector of the
      full theory.
\item Analogously to these factorization scales, characterized by the
      constants $C_{1}, C_{2}$, and $C_{3}$, we have introduced an
      additional arbitrary constant $C_{4}$ to define the
      renormalization scale $C_{4}f(x,y)Q = \mu _{\rm R}$, which appears
      in the argument of the running coupling
      $\alpha _{\rm s}^{\rm an}$
      (choice of renormalization prescription).
      This constant will play an important role in providing the link to
      the BLM (commensurate) scale-fixing.
      The running coupling plays a dual role: it describes the
      strength of the interaction at short distances (in the
      fixed-order perturbation theory), and controls via the anomalous
      dimensions of the cusped Wilson (world) line and the quark field,
      respectively, soft gluon emission and RG-evolution of
      $T_{\rm H}$ and ${\cal P}_{\pi}$ to the renormalization scale.
      The important point here is that the analytically improved
      running coupling contains a nonperturbative contribution, which
      reflects the nontrivial structure of the QCD physical vacuum.
\end{itemize}
The appropriate choice of the unphysical and arbitrary constants $C_{i}$
will be discussed in our numerical analysis in Sect.~\ref{sec:valid}.

%%%%%%%%%%%%%%%%%%%%%%%%%%%%%%%%%%%%%%%%%%%%%%%%%%%%%%%%%%%%%%%%%%%%%
%                            F I G U R E  2                         %
%                          \label{fig:qcdreg}                       %
%%%%%%%%%%%%%%%%%%%%%%%%%%%%%%%%%%%%%%%%%%%%%%%%%%%%%%%%%%%%%%%%%%%%%
%
%%%%%%%%%%%%%%%%%%%%%%%%%%%%%%%%%%%%%%%%%%%%%%%%%%%%%%%%%%%%%%%%%%%%%

The ambiguities parameterized by the scheme constants $C_{i}$ emerge
from the truncation of the perturbative series and would be absent
if one was able to derive all-order expressions in the coupling
constant.
In fact, the {\it calculated} (pion) form factor depends implicitly on
both scales: the adopted renormalization scale via $\alpha _{\rm s}$,
and the particular factorization scheme through the anomalous
dimensions.
Since the latter also depend on $\alpha _{\rm s}$, the
factorization-scheme and the renormalization-scheme dependences are
correlated.
On the other hand, the {\it physical} form factor is independent of
such artificial scales and satisfies
$
 \mu \frac{d F_{\pi}^{\rm phys} \left( Q^{2} \right)}{d\mu}
=
 0
$,
for $\mu$ being any internal scale.
Obviously, both scheme dependences should be treated simultaneously
and be minimized in order to improve the self-consistency of the
perturbative treatment.
In order to render the perturbative prediction reliable, the parameters
$C_{i}$ should be adjusted in such a way, as to minimize the influence
of higher-order corrections, thus resolving the scheme ambiguity.
However, in the present investigation we are not going to explicitly
match the fixed-order NLO contributions with the corresponding terms in
the re-summed expression for the ``soft'' logarithms.
Available calculations \cite{FGOC81,DR81,BT87,MNP98} of the NLO
contribution to the hard-scattering amplitude for the pion form factor
do not include the $k_{\perp}$ components of the gluon propagators,
making such a task difficult for the moment.
Instead, we are going to show in Sect.~\ref{sec:valid} that a potential
double counting of re-summed and NLO contributions is {\it de facto}
very small and of no real practical importance, especially in view of
the poor quality of the existing experimental data.
In addition, to limit a possible double counting as far as possible, we
have meticulously restricted the numerical evaluation of our analytic
expressions to the appropriate kinematical regimes.

In Eq.~(\ref{eq:piffbspace}), $T_{\rm H}$ is the amplitude for a quark
and an anti-quark to scatter via a series of hard-gluon exchanges with
gluonic transverse momenta (alias inter-quark transverse distances) not
neglected from the outset.
To leading order in the running coupling, one has
\begin{equation}
  T_{\rm H}\left(
                 x, y, b, Q; \mu _{\rm R}
           \right)
=
  8 C_{\rm F} \alpha _{\rm s}^{{\rm an}}(\mu _{\rm R}^{2})
  K_{0} \left( \sqrt{xy}\, bQ\right) \; .
\label{eq:T_Hbspace}
\end{equation}
%Eq (12) Hard-scattering amplitude for pion form factor in transverse
%        configuration space at LO
This result is related to the more familiar momentum-space expression
\begin{equation}
  T_{\rm H}
        \left(
              x, y, \kv ,\lv ,Q, \mu _{\rm R}
        \right)
=
    \frac{16 \pi \, C_{\rm F} \,
    \alpha _{\rm s}(\mu _{\rm R}^{2})}{xyQ^{2}
  + \left( \kv + \lv \right)^{2}}
\label{eq:T_Hkspace}
\end{equation}
%Eq (13) Hard scattering amplitude for pion form factor depending on
%        transverse momenta at LO
via the Fourier transformation
\begin{equation}
  T_{\rm H}
           \left(x, y, \kv , \lv , Q, \mu _{\rm R}
           \right)
=
  \int_{-\infty}^{\infty} d{}^{2}\bv \,
  T_{\rm H}\left( x, y, b, Q, \mu _{R} \right)
  \exp \left[
             i \bv \cdot \left(
                                \kv + \lv
                         \right)
       \right] \; ,
\label{eq:fourierT_H}
\end{equation}
%Eq (14) Fourier transformation of hard scattering amplitude T_H
where use of the symmetry of $\psi _{\pi}$ under
$x \leftrightarrow 1-x \equiv \bar{x}$ has been made, and where
$C_{\rm F}=(N_{\rm c}^{2} - 1)/2N_{\rm c} = 4/3$ for $SU(3)_{\rm c}$.
In the limit $Q^{2}\to \infty$ (and $x$ fixed) this expression coincides
with the asymptotic hard scattering in the collinear approximation, up
to suppressed power corrections.
The latter become important in the end-point region ($x\to 0$) at fixed
$Q^{2}$, where the actual momentum flow in the gluon propagator becomes
small (typically of the order of $\Lambda _{\rm QCD}$) so that gluonic
transverse momenta cannot be safely neglected.

The amplitude
\begin{eqnarray}
\everymath{\displaystyle}
  {\cal P}_{\pi}\left( x, b, P\simeq Q, C_{1}, C_{2}, \mu
                \right)
=
 \exp \Biggl[
             && -  s\left( x, b, Q, C_{1}, C_{2} \right)
                -  s\left( \bar{x}, b, Q, C_{1}, C_{2} \right)
\nonumber \\
&&
                - 2 \int_{C_{1}/b}^{\mu} \frac{d\bar\mu}{\bar\mu} \,
                \gamma _{q}\left(\alpha _{\rm s}^{\rm an}(\bar\mu )
                           \right)
      \Biggr]
  {\cal P}_{\pi}\left( x, b, C_{1}/b \right)
\label{eq:piamplbspace}
\end{eqnarray}
%Eq (15) Modified pion distribution amplitude including Sudakov
%        corrections and intrinsic k_perp distribution
describes the distribution of longitudinal momentum fractions of the
q\=q pair, taking into account the intrinsic transverse size of the
pion state \cite{JK93} and comprising corrections due to soft real and
virtual gluons \cite{LS92}, including also evolution from the initial
amplitude ${\cal P}_{\pi}\left( x, b, C_{1}/b \right)$ at point
$C_{1}/b$ to the renormalization scale $\mu \propto Q$.
Let us emphasize at this point that the power-behaved term,
$\alpha _{\rm}^{\rm npert}$, does not change the leading double
logarithmic behavior of the Sudakov exponent.
The main effect of the absence of a Landau pole in the running coupling
$\alpha _{\rm s}^{\rm an}$ is to make the functions
$s\left( x, b, Q, C_{1}, C_{2} \right)$,
$s\left( \bar{x}, b, Q, C_{1}, C_{2} \right)$
well-defined (analytic) in the IR region and to slow down evolution by
extending soft-gluon cancellation down to the scale
$C_{1}/b \simeq \Lambda _{\rm QCD}$, where the full Sudakov form factor
acquires a finite value, modulo its $Q^{2}$ dependence (see
Fig.~\ref{fig:sudalepj}).
In addition, as we shall see below, the Sudakov exponent contains
power-behaved corrections in
$\left(C_{1}/b\Lambda\right)^{2p}$ and
$\left(C_{2}/\xi Q\Lambda\right)^{2p}$, starting with $p=1$.
Such contributions are the footprints of soft gluon emission at the
kinematic boundaries to the non-perturbative QCD regime, characterized
by the transversal (or IR) and the longitudinal (or collinear) cutoffs.

The pion distribution amplitude evaluated at the (low) factorization
scale $C_{1}/b$ is approximately given by
\begin{equation}
  {\cal P}_{\pi}\left( x, b, C_{1}/b, m_{\rm q} \right)
\simeq
  \frac{f_{\pi}/\sqrt{2}}{2\sqrt{N_{\rm c}}} \,
  \phi _{\pi}\left( x, C_{1}/b \right)
  \Sigma \left( x, b, m_{\rm q} \right) \; .
\label{eq:inpiampl}
\end{equation}
%Eq (16) Pion distribution amplitude at low factorization scale
Because we retain the intrinsic transverse momenta of the pion bound
state, we have to make an ansatz for their distribution.
In the present work, we follow Brodsky, Huang, and Lepage \cite{BHL83}
(see, also \cite{JK93,JKR96}) and parameterize the distribution
$\Sigma (x, {\kv}, m_{\rm q})$
in the intrinsic transverse momentum $k_{\perp}$
(or equivalently the intrinsic inter-quark transverse distance $b$)
in the form of a {\it non-factorizing} in the variables $x$ and
$k_{\perp}$ (or $x$ and $b$) Gaussian function, so that
\begin{equation}
  \Psi _{\pi} \left(x, {\kv}, C_{1}/b, m_{\rm q}\right)
=
  \frac{f_{\pi}/\sqrt{2}}{2N_{\rm c}}
  \Phi _{\pi}(x, C_{1}/b)
  \Sigma \left(x, {\kv}, m_{\rm q}\right) \; ,
\label{eq:psiansatz}
\end{equation}
%Eq (17) Ansatz for pion wave function
where
\begin{equation}
  \Phi _{\pi}\left( x, C_{1}/b \right)
=
  A\, \Phi _{\rm as}(x)
=
  A\, 6x(1-x)
\label{eq:asymptoticwf}
\end{equation}
%Eq (18) Asymptotic distribution amplitude
is the asymptotic distribution amplitude, with $A$ being an appropriate
normalization factor, and where
\begin{equation}
  \Sigma \left(x, {\kv}, m_{\rm q}\right)
=
  16 \pi ^{2} \beta ^{2} g(x)
  \hat\Sigma \left(x, {\kv}\right)
  \hat\Sigma \left(x, m_{\rm q}\right)
\label{eq:FullGaussiankspace}
\end{equation}
%Eq (19) Gaussian ansatz for k_perp distribution with quark mass
with
\begin{equation}
  \hat\Sigma \left(x, {\kv}\right)
=
  \exp{\left[-\beta ^{2} {\kv}^{2} g(x)\right]} \; ,
\label{eq:k_perpGaussian}
\end{equation}
%Eq (20) Gaussian ansatz for intrinsic transverse momentum
and
\begin{equation}
  \hat\Sigma \left(x, m_{\rm q}\right)
=
  \exp{\left[-\beta ^{2} m_{\rm q}^{2} g(x)\right]} \; ,
\label{eq:massGaussian}
\end{equation}
%Eq (21) Mass term in Gaussian ansatz
models the distribution in the intrinsic transverse momentum in the form
of a Gaussian in the sense of the BHL ansatz.\footnote{The width
parameter $\beta$ of the Gaussian distribution should not be confused
with the $\beta$-function.}

Neglecting transverse momenta in Eq.~(\ref{eq:T_Hkspace}) (collinear
approximation), the only dependence on $k_{\perp}$ resides in the wave
function.
Furthermore, limiting the maximum value of $k_{\perp}$, these degrees of
freedom can be integrated out independently for the initial and final
pion states to give way to the corresponding pion distribution
amplitudes, which depend only implicitly on the cutoff momentum
$\mu ^{2}=\left(C_{1}/b\right)^{2}$:
\begin{equation}
  \frac{f_{\pi}/\sqrt{2}}{2\sqrt{N_{\rm c}}} \,
  \phi _{\pi}\left( x, C_{1}/b \right)
=
  \int_{}^{\left(C_{1}/b\right)^{2}} \frac{d^{2}\kv}{16 \pi ^{3}} \,
  \Psi _{\pi} \left( x, {\kv}, m_{\rm q} \right) \; ,
\label{eq:pida}
\end{equation}
%Eq (22) Pion distribution amplitude
where
$f_{\pi} = 130.7$~MeV and $N_{\rm c}=3$.
Setting $g(x)=1/\left(x(1-x)\right)$ and integrating on both sides of
this equation over $x$, supplies us with the constraint
\begin{equation}
  1= A \int_{0}^{1} dx 6 x (1-x)
  \exp{\left[-\frac{\beta ^{2} m_{\rm q}^{2}}{x(1-x)}\right]}
\label{eq:phinormalization}
\end{equation}
%Eq (23) Phi normalization
because by virtue of the leptonic decay
$\pi \rightarrow \mu ^{+} \nu _{\mu}$
the rhs of Eq.~(\ref{eq:pida}) is fixed to
$
 \frac{f_{\pi}/\sqrt{2}}{2\sqrt{N_{\rm c}}}
$,
so that
\begin{equation}
  \int_{0}^{1}dx \, \phi _{\pi}\left( x, \mu ^{2}, m_{\rm q}=0 \right)
=
  1
\label{eq:normphi}
\end{equation}
%Eq (24) Normalization to unity of Phi_pi
and
\begin{equation}
  \int \frac{d{}^{2}{\kv}}{16\pi^{3}}
  \hat\Sigma \left(x, {\kv}\right)
=
  1
\label{eq:k_normal}
\end{equation}
%Eq (25) Normalization of Gaussian ansatz in k_perp
for any factorization (normalization) scale $\mu$.
The contributions from higher Fock states are of higher twist and
contribute corrections at higher order in $1/Q$, i.e., they are
power-law suppressed.

Moreover, from $\pi ^{0} \to \gamma\gamma$, we have
\begin{equation}
  1
=
  8 A f_{\pi}^{2} \pi ^{2} \beta ^{2}
  \int_{0}^{1} dx
  \exp{\left[-\frac{\beta ^{2} m_{\rm q}^{2}}{x(1-x)}\right]} \; ,
\label{eq:normalization}
\end{equation}
%Eq (26) Normalization
while the quark probability $P_{\rm{q\bar q}}$ and the average
transverse momentum $\langle {\kv}^{2}\rangle$ are given by
\begin{equation}
  P_{\rm{q\bar q}}
=
  12 A^{2} f_{\pi}^{2} \pi ^{2} \beta ^{2}
  \int_{0}^{1} dx x(1-x)
  \exp{\left[-\frac{2\beta ^{2} m_{\rm q}^{2}}{x(1-x)}\right]} \; ,
\label{eq:probability}
\end{equation}
%Eq (27) Probability for valence state
\begin{equation}
  \langle {\kv}^{2}\rangle
=
  \frac{6 A^{2} f_{\pi}^{2} \pi ^{2}}{P_{\rm{q\bar q}}}
  \int_{0}^{1} dx x^{2}(1-x)^{2}
  \exp{\left[-\frac{2\beta ^{2} m_{\rm q}^{2}}{x(1-x)}\right]} \; .
\label{eq:expectationkperp}
\end{equation}
%Eq (28) Expectation value of k_perp
By inputting $f_{\pi}$ and the value of the quark mass $m_{\rm q}$,
we determine the parameters
$A$, $\beta ^{2}$, $P_{\rm{q\bar q}}$, and
${\langle {\kv}^{2}\rangle}^{1/2}$ (see, Table \ref{tab:parameters}).

%%%%%%%%%%%%%%%%%%%%%%%%%%%%%%%%%%%%%%%%%%%%%%%%%%%%%%%%%%%%%%%%%%%%%
%                           T A B L E  1                            %
%                      label{tab:parameters}                        %
%%%%%%%%%%%%%%%%%%%%%%%%%%%%%%%%%%%%%%%%%%%%%%%%%%%%%%%%%%%%%%%%%%%%%
\begin{table}
\caption[tab:wfparas]
        {Values of parameters entering the pion wave function, using
         the notations of \cite{JK93}.
         The values in parentheses refer to the case $m_{\rm q}=0$.
         Here the subscript as on $\beta ^{2}$ refers to the asymptotic
         distribution amplitude.
\label{tab:parameters}}
\begin{tabular}{lc}
Input parameters & Determined parameters \\
\hline
$m_{\rm q}=0.33$~GeV & $A=\frac{1}{6}\cdot 10.01$
$(\frac{1}{6}\cdot 6)$\\
$f_{\pi}=0.1307$~GeV & $\beta _{\rm as}^{2}=0.871$~GeV${}^{-2}$
$(0.743$~GeV${}^{-2})$\\
& ${\langle {\kv}^{2}\rangle}^{1/2}=0.352$~GeV
$(0.367$~GeV)\\
& $P_{\rm{q\bar q}}=0.306$
$(0.250)$\\
\end{tabular}
\end{table}
%T A B L E  1
%
%%%%%%%%%%%%%%%%%%%%%%%%%%%%%%%%%%%%%%%%%%%%%%%%%%%%%%%%%%%%%%%%%%%%%

How large should be the quark mass used in the BHL ansatz?
The parameter $m_{\rm q}$ has not the meaning of a real mechanical
mass for the quark, but reflects the complicated structure of the QCD
vacuum.
Let us make this point more transparent.
The QCD Lagrangian contains no mass scale in the chiral limit.
A mass scale enters perturbatively only through dimensional
transmutation to enable renormalization.
Nonperturbative scales derive from vacuum fluctuations of some definite
correlation length in the context of specified QCD vacuum models.
For instance, in the chiral quark model derived from the instanton
approach (\cite{DP86,DPP88}), the pivotal nonperturbative scale is the
average instanton size $\rho$, whose inverse defines a mass scale of
the order of $0.600$~GeV.
Breaking chiral symmetry spontaneously in the instanton vacuum by the
delocalization of the fermionic zero modes, the massless quark acquires
a momentum-dependent mass to become a quasi-particle.
The obtained value of this effective mass in the quark propagator is
$M \sim \rho /R^{2} \simeq 0.300-0.350$~GeV, where $R$ denotes
the separation between the quark and the antiquark.
Note that though one deals with massive quarks, higher Fock states are
not necessarily zero, as the quark propagator contains a
renormalization factor
$Z \sim 1 + O(\rho ^{2}/R^{2})$
from which one infers that parton quarks and the ``constituent'' quarks
(the quasi-particles) of this model are equal at leading order in the
small parameter
$\rho ^{2}/R^{2}$.
Hence, on the basis of the nonperturbative structure of the QCD vacuum,
we may conclude that the mass scale characterizing the quarks in the
pion is of the order of the typical constituent quark mass.
Then, it seems plausible to set the mass scale $m_{\rm q}$ in the BHL
ansatz equal to the dynamical mass $M_{\rm q}$ obtained in such a
nonperturbative vacuum model, rather than to use a current quark mass
of a few MeV.
Only for a very dilute instanton vacuum one can realize
$M_{\rm q}\ll\rho ^{-1}$, but then the shape of the pion distribution
amplitude deviates significantly from the asymptotic one becoming very
flat.
Actually, one may considerably simplify the BHL ansatz by setting all
scales responsible for the intrinsic (transverse) structure of the
hadron (the pion), namely
${\langle {\kv}^{2}\rangle}^{1/2}$
and
$m_{\rm q}$
equal to
$\Lambda _{\rm QCD}$
which is of the same order of magnitude:
$0.200 - 0.350$~GeV.\footnote{The average transverse momentum of the
pion was recently determined in \cite{BRGS00} on the basis of local
duality. Values between $0.260$ and $0.320$ GeV were found, obviously
consistent with the actual value of $\Lambda _{\rm QCD}$ and those in
Table \ref{tab:parameters}.}
Indeed, the predictions obtained with this simplification are close to
those calculated with the values given in Table \ref{tab:parameters}.

Let us now return to Eq.~(\ref{eq:piamplbspace}).
The Sudakov form factor
$F_{\rm S}\left(\xi , b, Q, C_{1}, C_{2}\right)$,
i.e., the exponential factor in front of the wave function, will be
expressed as the expectation value of an open Wilson (world) line along
a contour of finite extent, $C$, which follows the bent quark line in
the hard-scattering process from the segment with direction $P$ to that
with direction $P^{\prime}$ after being abruptly derailed by the hard
interaction which creates a ``cusp'' in $C$, and is to be evaluated
within the range of momenta termed ``soft'', confined within the range
limited by $C_{1}/b$ (IR cutoff) and $C_{2}\xi Q$ (longitudinal cutoff)
(where $\xi = x, \bar{x}, y, \bar{y}$).\footnote{This means that the
region of hard interaction of the Wilson line with the off-shell photon
is factorized out.}
Thus we have \cite{Col89,KR87,Kor89,GKKS97,KR92}
\begin{equation}
  F_{\rm S}\left(W(C)\right)
=
  \left\langle {\rm P}
  \exp \left(
             i g \int_{C}^{} dz
             \cdot t^{a}
             A^{a}(z)
       \right)
  \right \rangle _{\rm soft} \; ,
\label{eq:wilson}
\end{equation}
%Eq (29) Sudakov form factor as expectation value of Wilson line
where {\rm P} stands for path ordering along the integration contour
$C$, and where ${<...>}_{A}$ denotes functional averaging in the gauge
field sector with whatever this may entail (ghosts, gauge choice
prescription, Dirac determinant, etc.).
Having isolated a sub-sector of the full theory
(cf.~Fig.~\ref{fig:qcdreg}), where only gluons with virtualities
between $C_{1}/b$ and $C_{2}\xi Q$ are active degrees of freedom,
quark propagation and gluon emission can be described by eikonal
techniques, using either Feynman diagrams \cite{CS81,BS89} or by
employing a world-line casting of QCD which reverts the fermion
functional integral into a first-quantized, i.e., particle-based
path integral \cite{GKKS97}.

Then the Sudakov functions, entering Eq.~(\ref{eq:piamplbspace}), can be
expressed in terms of the momentum-dependent cusp anomalous dimension
of the bent contour to read
\begin{equation}
  s\left(\xi , b, Q, C_{1}, C_{2} \right)
=
  \frac{1}{2}
  \int_{C_{1}/b}^{C_{2}\xi Q}
  \frac{d\mu}{\mu} \,
  \Gamma _{{\rm cusp}}
        \left(\gamma , \alpha _{\rm s}^{\rm an}(\mu )
        \right)
\label{eq:sudfuncusp}
\end{equation}
%Eq (30) Sudakov function in terms of cusp anomalous dimension
with the anomalous dimension of the cusp given by
\begin{eqnarray}
  \Gamma _{{\rm cusp}}
        \left( \gamma , \alpha _{\rm s}^{\rm an}(\mu ) \right)
& = &
  2 \ln \left(
              \frac{C_{2}\xi Q}{\mu}
        \right) A\left( \alpha _{\rm s}^{\rm an}(\mu ) \right)
              + B\left( \alpha _{\rm s}^{\rm an}(\mu ) \right) \; ,
\nonumber \\
& \equiv &
    \Gamma _{{\rm cusp}}^{\rm pert}
  +
    \Gamma _{{\rm cusp}}^{\rm npert} \; ,
\label{eq:gammacusp}
\end{eqnarray}
%Eq (31) Sudakov function in NLO
$\gamma = \ln \left(C_{2}\xi Q/\mu \right)$
being the cusp angle, i.e., the emission angle of a soft gluon and the
bent eikonalized quark line after the external (large) momentum $Q$ has
been injected at the cusp point by the off-mass-shell photon, and where
in the second line of Eq.~(\ref{eq:gammacusp}) the superscripts relate
to the origin of the corresponding terms in the running coupling
(cf.~Eqs.~(\ref{eq:oneloopalpha_an}), (\ref{eq:twoloopalpha_an})).
The functions $A$ and $B$ are known at two-loop order:
\begin{eqnarray}
\everymath{\displaystyle}
  A\left( \alpha _{\rm s}^{\rm an}(\mu ) \right)
& = &
  \frac{1}{2}
  \left[
         \gamma _{\cal K}
                 \left( \alpha _{\rm s}^{\rm an}(\mu ) \right)
       + \beta (g) \frac{\partial}{\partial g}
       {\cal K}(C_{1}, \alpha _{\rm s}^{\rm an}(\mu ))
  \right]
\nonumber \\
& = &
       C_{\rm F} \frac{\alpha _{\rm s}^{\rm an}(g(\mu ))}{\pi}
     + \frac{1}{2} K\left( C_{1} \right) C_{\rm F}
       \left( \frac{\alpha _{\rm s}^{\rm an}(g(\mu ))}{\pi} \right)^{2}
\; ,
\label{eq:funAnlo}
\end{eqnarray}
%Eq (32) Function A in NLO
and
\begin{eqnarray}
\everymath{\displaystyle}
  B\left( \alpha _{\rm s}^{\rm an}(\mu ) \right)
& = &
  - \frac{1}{2}
    \left[
          {\cal K} \left(
                         C_{1}, \alpha _{\rm s}^{\rm an}(\mu )
                   \right)
        + {\cal G} \left(
                         \xi, C_{2}, \alpha _{\rm s}^{\rm an}(\mu )
                   \right)
    \right]
\nonumber \\
& = &
      \frac{2}{3} \frac{\alpha _{\rm s}^{\rm an}(g(\mu ))}{\pi}
      \ln \left(
          \frac{C_{1}^{2}}{C_{2}^{2}}\frac{{\rm e}^{2\gamma_{E}-1}}{4}
          \right) \; .
\label{eq:funBnlo}
\end{eqnarray}
%Eq (33) Function B in NLO
The first term in Eq.~(\ref{eq:funAnlo}) is universal,\footnote{In works
quoted above, the cusp anomalous dimension is identified with the
universal term, whereas the other (scheme and/or process dependent)
terms are considered as additional anomalous dimensions.
Here this distinction is irrelevant.}
while the second one as well as the contribution termed $B$ are scheme
dependent.
The K-factor in the ${\overline{\rm MS}}$ scheme to two-loop order
is given by \cite{Col89,KR87,CS81,DS84,KT82}
\begin{equation}
  K\left(C_{1}\right)
=
     \left(\frac{67}{18} - \frac{\pi ^{2}}{6}\right) C_{\rm A}
   - \frac{10}{9}n_{\rm f} T_{\rm F}
   + \beta _{0} \ln \left( C_{1} {\rm e}^{\gamma _{\rm E}}/2 \right)
\label{eq:Kfactor}
\end{equation}
%Eq (34) K-factor in \bar{MS} scheme
with $C_{\rm A}=N_{\rm C}=3$, $n_{\rm f}=3$, $T_{\rm F}=1/2$,
and $\gamma _{\rm E}$ being the Euler-Mascheroni constant.

The quantities ${\cal K}$, ${\cal G}$ in Eq.~(\ref{eq:funBnlo}) are
calculable using the non-Abelian extension to QCD \cite{CS81} of the
Grammer-Yennie method \cite{GY73} for QED.
Alternatively, one can calculate the cusp anomalous dimension employing
Wilson (world) lines \cite{KR87,Kor89,GKKS97,KR92}.\footnote{The
derivation of the cusp anomalous dimension in the $1/N_{\rm f}$
approximation (single-bubble-chain approximation) was given in
\cite{BB95}, Appendix A.}
In this latter approach (see, e.g., \cite{GKKS97}), the IR behavior of
the cusped Wilson (world) line is expressed in terms of an effective
fermion vertex function whose variance with the momentum scale is
governed by the anomalous dimension of the cusp within the isolated
effective sub-sector (see Fig.~\ref{fig:qcdreg}).
Since this scale dependence is entirely restricted within the
low-momentum sector of the full theory, IR scales are locally coupled
and the soft (Sudakov-type) form factor depends only on the cusp angle
which varies with the inter-quark transverse distance $b$ ranging
between $C_{1}/b$ and $C_{2}\xi Q$.

The corresponding anomalous dimensions are linked to each other (for a
nice discussion, see \cite{Col89}) through the relation
$
  2 \Gamma _{{\rm cusp}} \left( \alpha _{\rm s}^{\rm an}(\mu ) \right)
=
  \gamma _{\cal K}\left( \alpha _{\rm s}^{\rm an}(\mu ) \right)
$
with
$
 \Gamma _{{\rm cusp}} (\alpha _{\rm s}^{\rm an}(\mu ))
=
  C_{\rm F}\, \alpha _{\rm s}^{\rm an}(\mu ^{2})/\pi
$, which shows that
$
 \frac{1}{2}\gamma _{\cal K}
=
 A\left( \alpha _{\rm s}^{\rm an}(\mu ) \right)
$.
(Note that $\gamma _{\cal G} = - \gamma _{\cal K}$.)

The soft amplitude
${\cal P}_{\pi}\left( x, b, C_{1}/b, \mu \right)$
and the hard-scattering amplitude
$T_{\rm H}\left( x, y, b, Q, \mu \right)$
satisfy independent RG equations to account for the dynamical
factorization (recall that both $b$ and $\xi$ are integration variables)
with solutions controlled by the power-term modified ``evolution time''
(see, e.g., \cite{DDT80} and earlier references cited therein):
\begin{eqnarray}
\everymath{\displaystyle}
  \tau \left( \frac{C_{1}}{b}, \mu \right)
& = &
  \int_{C_{1}^{2}/b^{2}}^{\mu ^{2}} \frac{dk^{2}}{k^{2}} \,
  \frac{\alpha _{\rm s}^{{\rm an}(1)}(k^{2})}{4\pi}
\nonumber \\
& = &
  \frac{1}{\beta _{0}}
  \ln \frac{\ln\left(\mu ^{2}/\Lambda ^{2}\right)}
           {\ln\left(C_{1}^{2}/\left( b\Lambda \right)^{2}
               \right)}
+ \frac{1}{\beta _{0}}
  \left[
  \ln \frac{\mu ^{2}}{\left( C_{1}/b \right)^{2}}
- \ln \frac{\left\vert\mu ^{2} - \Lambda ^{2}\right\vert}
           {\left\vert\frac{C_{1}^{2}}{b^{2}} - \Lambda ^{2}\right\vert}
  \right]
\label{eq:evoltime}
\end{eqnarray}
%Eq (35) Modified evolution time
from the factorization scale
$C_{1}/b$ to the observation scale $\mu$, with $\Lambda$
denoting $\Lambda _{\rm QCD}$ as before.
The evolution time is directly related to the quark anomalous dimension,
viz.,
$
 \gamma _{\rm q}\left( \alpha _{\rm s}^{\rm an}(\mu ) \right)
=
 - \alpha _{\rm s}^{{\rm an}}(\mu ^{2})/\pi
$.
One appreciates that the second term in (\ref{eq:evoltime}) stems
from the power-generated correction to the running coupling,
$\alpha _{\rm s}^{\rm npert}$, and is absent in the conventional
approach.
At moderate values of $\mu ^{2}$ this term is ``slowing down'' the rate
of evolution.

The leading contribution to the IR-modified Sudakov functions
$s\left( \xi ,b, Q, C_{1}, C_{2} \right)$
(where $\xi = x, \bar{x}, y, \bar{y}$)
is obtained by expanding the functions $A$ and $B$ in a power
series in $\alpha _{\rm s}^{\rm an}$ and collecting together all
large logarithms
$
 \left( \frac{\alpha _{\rm s}^{\rm an}}{\pi} \right)^{n}
 \ln \left ( \frac{C_{2}}{C_{1}} \xi b Q \right)^{m}
$,
which can be transformed back into large logarithms
$
 \ln \left( Q^{2}/{\kv}^{2} \right)
$
in transverse momentum space.
Employing equations (\ref{eq:oneloopalpha_an}) and
(\ref{eq:twoloopalpha_an}), the leading contribution results from the
expression
\begin{eqnarray}
\everymath{\displaystyle}
  s\left( \xi , b, Q, C_{1}, C_{2}\right)
= &&
    \frac{1}{2}
    \int_{C_{1}/b}^{C_{2}\xi Q} \frac{d\mu}{\mu}
    \Biggl\{
           2 \ln \left( \frac{C_{2} \xi Q}{\mu} \right)
       \Biggl[
         \frac{\alpha _{\rm s}^{{\rm an}(2)}(\mu )}{\pi}
         A^{(1)}
      +  \left(
         \frac{\alpha _{\rm s}^{{\rm an}(1)}(\mu )}{\pi}
         \right)^{2}
         A^{(2)}\left( C_{1} \right)
       \Biggr]
\nonumber \\
&&    +  \frac{\alpha _{\rm s}^{{\rm an}(1)}(\mu )}{\pi}
         B^{(1)}\left( C_{1}, C_{2} \right)
      +  {\cal O} \left( \frac{\alpha _{\rm s}^{\rm an}}{\pi}\right)^{3}
    \Biggr\} \; ,
\label{eq:SudakovNLO}
\end{eqnarray}
%Eq (36) Sudakov at NLO in IR-finite scheme
where Eq.~(\ref{eq:twoloopalpha_an}) is to be used in front of
$A^{(1)}$, whereas the other two terms are to be evaluated with
Eq.~(\ref{eq:oneloopalpha_an}).
The specific values of the coefficients $A^{(i)}$, $B^{(i)}$ are
\begin{eqnarray}
\everymath{displaystyle}
  A^{(1)}                         & = & C_{\rm F} \; ,
\nonumber \\
  A^{(2)} \left(C_{1}\right)      & = & \frac{1}{2}\, C_{\rm F} \,
                                        K\left( C_{1} \right) \; ,
\nonumber \\
  B^{(1)}\left(C_{1},C_{2}\right) & = & \frac{2}{3}
                \ln \left(
                          \frac{C_{1}^{2}}{C_{2}^{2}}
                          \frac{{\rm e}^{2\gamma_{E}-1}}{4}
                    \right) \; ,
\label{eq:expcoef}
\end{eqnarray}
%Eq (37) Expansion coefficients for the Sudakov function up to NLLO
in which the term proportional to $A^{(1)}$ represents the universal
part.
As now the power-correction term in $\alpha _{\rm s}^{\rm an}$ gives
rise to poly-logarithms, a formal analytic expression for the full
Sudakov form factor is too complicated for being presented.
We only display the universal contribution in LLA:
\begin{eqnarray}
\everymath{\displaystyle}
  F_{\rm S}^{\rm univ}\left(\mu _{\rm F}, Q\right)
=
&&
  \exp \Biggl\{
               - \frac{C_{\rm F}}{\beta _{0}}
       \Biggl[
                 \ln \left(\frac{\tilde{Q}^{2}}{\Lambda ^{2}}\right)
                 \ln \frac{\ln \tilde{Q}^{2}/\Lambda ^{2}}
                          {\ln \mu _{\rm F}^{2}/\Lambda ^{2}}
               - \ln\frac{\tilde{Q}^{2}}{\mu _{\rm F}^{2}}
\nonumber \\
&&
  + \ln\left(\frac{\tilde{Q}^{2}}{\mu _{\rm F}^{2}}\right)
    \ln\frac{{\Lambda}^{2}-\mu _{\rm F}^{2}}{\Lambda ^{2}}
  + \frac{1}{2}\, \ln ^{2} \frac{\tilde{Q}^{2}}{\mu _{\rm F}^{2}}
  + {\rm Li}_{2}\left(\frac{\tilde{Q}^{2}}{\Lambda ^{2}}\right)
  - {\rm Li}_{2}\left(\frac{\mu _{\rm F}^{2}}{\Lambda ^{2}}\right)
       \Biggr]
       \Biggl\} \; ,
\label{eq:unipart}
\end{eqnarray}
%Eq (38) Universal part in LLA of IR-modified Sudakov form factor
where $\tilde{Q}$ represents the scale $C_{2}\xi Q$ and the IR matching
(factorization) scale $\mu _{\rm F}$ varies with the inverse transverse
distance $b$, i.e., $\mu _{\rm F} = C_{1}/b$.
Note that the four last terms in this equation originate from the
non-perturbative power correction (cf.~Eq.~(\ref{eq:gammacusp})), and
that ${\rm Li}_{2}$ is the dilogarithm (Spence) function which comprises
power-behaved corrections of the IR ($b\Lambda$) and the longitudinal
($Q/\Lambda$) cutoff scales.
In the calculations to follow, Eq.~(\ref{eq:SudakovNLO}) is evaluated
numerically to NLLA with appropriate kinematic bounds to ensure
proper factorization at the numerical level.
Above (see Eq.~(\ref{eq:SudakovNLO})), we have replaced
$\left(\alpha _{\rm s}^{(1)}\right)^{2}$ by
$\left(\alpha _{\rm s}^{{\rm an}(1)}\right)^{2}$.
Here we have an analytization ambiguity.
Since nonlinear relations are not preserved by the analytization
procedure \cite{Shi99} (see, also \cite{BRS00}),
we could have made the square of the running coupling,
$
 \left(\alpha _{\rm s}^{(1)}\right)^{2}
$,
analytic as a whole.
We plan to report on these interesting conceptual issues of
analytization in a separate publication.

Note that, neglecting the power-generated logarithms, we obtain an
equation for the conventional Sudakov function, which we write as an
expansion in inverse powers of the first beta-function coefficient
$\beta _{0}$ to read
\begin{eqnarray}
\everymath{\displaystyle}
  s\left( \xi , b, Q, C_{1}, C_{2} \right)
= &&
  \frac{1}{\beta _{0}}
  \left[
         \left(
               2 A^{(1)} \hat Q + B^{(1)}
         \right)\ln \frac{\hat Q}{\hat b}
       - 2 A^{(1)} \left( \hat Q - \hat b \right)
  \right]
\nonumber \\
&& - \, \frac{4}{\beta _{0}^{2}}
     A^{(2)}
     \left(
           \ln \frac{\hat Q}{\hat b} -
           \frac{\hat{Q} - \hat{b}}{\hat b}
     \right)
\nonumber \\
&& + \frac{\beta _{1}}{\beta _{0}^{3}} A^{(1)}
     \left\{
              \ln \frac{\hat Q}{\hat b}
            - \frac{\hat Q - \hat b}{\hat b}
              \left[
                    1 + \ln \left( 2 \hat b \right)
              \right]
            + \frac{1}{2}
              \left[
                     \ln ^{2}\left(2\hat Q \right)
                   - \ln ^{2}\left(2\hat b \right)
              \right]
     \right\} \; ,
\label{eq:oldsudakov}
\end{eqnarray}
%Eq (39) Correct expression for conventional Sudakov function
where the convenient abbreviations \cite{LS92}
$
 \hat Q
\equiv
 \ln \frac{C_{2} \xi Q}{\Lambda}
$
and
$
 \hat b
\equiv
  \ln \frac{C_{1}}{b\Lambda}
$
have been used.

This quantity differs from the original result given by Li and Sterman
in \cite{LS92}, and, though it almost coincides numerically with the
formula derived by J. Bolz \cite{Bol95}, it differs from that
algebraically.

All told, the final expression for the electromagnetic pion form factor
at leading perturbative order in $T_{\rm H}$ and next-to-leading
logarithmic order in the Sudakov form factor has the form
\begin{eqnarray}
\everymath{\displaystyle}
  F_{\pi}(Q^{2})
= &&
  \frac{2}{3}\, A^{2} \pi\, C_{\rm F}\, f_{\pi}^{2}\,
  \int_{0}^{1} dx
  \int_{0}^{1} dy
  \int_{0}^{\infty}b\, db\,
  \alpha _{\rm s}^{{\rm an}(1)}\left(\mu _{\rm R}\right)
  \Phi _{\rm as} (x)
  \Phi _{\rm as} (y)
  \exp \left[-\frac{b^{2} \left(x\bar{x} + y\bar{y}\right)}
  {4\beta _{\rm as}^{2}}\right]
\nonumber \\
&& \times
  \exp{ \left[-\beta _{\rm as}^{2} m_{\rm q}^{2}
  \left(\frac{1}{x\bar x} + \frac{1}{y\bar y}\right)
        \right]}
  K_{0}\left(\sqrt{xy}Qb\right)
   \exp \left[ - S \left(x,y,b,Q, C_{1},C_{2},C_{4}\right) \right] \; ,
\label{eq:pifofafin}
\end{eqnarray}
%Eq (40) Final form for electromagnetic form factor of pion
where
\begin{equation}
  S \left(x,y,b,Q, C_{1},C_{2},C_{4}\right)
\equiv
   s\left(x, b, Q, C_{1}, C_{2}\right)
 + s\left(\bar{x}, b, Q, C_{1}, C_{2}\right)
 + (x \leftrightarrow y)
 - 8 \,\tau \left(\frac{C_{1}}{b}, \mu _{\rm R}\right)
\label{eq:fullSudakov}
\end{equation}
%Eq (41) Full Sudakov expression, i.e., including evolution
with $\tau \left(C_{1}/b, \mu _{\rm R}\right)$ given by
Eq.~(\ref{eq:evoltime}) and $\mu _{\rm R} = C_{4}f(x,y)Q$.
As we shall show below, the effect of including the effective quark
mass in the hard part of the form factor is almost negligible, as one
should expect on theoretical grounds.

Before we go beyond the leading order in the perturbative expansion of
the hard-scattering amplitude, $T_{\rm H}$, let us pause for a moment
to comment on the pion wave function.
We have pro-actively indicated in Eq.~(\ref{eq:pifofafin}) that the
asymptotic distribution amplitude
$\Phi _{\rm as}(x) = 6 x \bar x$
will be used.

A few words about this choice are now in order.

Hadron wave functions are clearly the essential variables needed to
model and describe the properties of an intact hadron.
In the past, most attempts to improve the theoretical predictions for
the hard contribution to the pion form factor have consisted of using
end-point concentrated wave functions (distribution amplitudes).
In this analysis we refrain from using such distribution amplitudes of
the Chernyak-Zhitnitsky (CZ) type \cite{CZ84}, referring for a
compilation of objections and references to \cite{SSK98} (see also
\cite{Rad98}), and present instead evidence for an alternative source of
enhancement due to the non-perturbative power correction in the running
coupling.

This IR-enhancement effect was found in \cite{SSK98} to be quite
significant, even for the asymptotic solution (to the evolution
equation) which has its maximum at $x=1/2$.
Indeed, the IR-enhanced hard contribution can account already at
leading perturbative order for a sizable part of the measured magnitude
of the electromagnetic pion form factor, though agreement with the
currently available experimental (low-momentum) data calls for the
inclusion of the soft, non-factorizing contribution
(cf.~Eq.~(\ref{eq:softhard})) \cite{IL84,Rad90,BH94,JKR96} -- even if
the NLO correction is taken into account (see Sect.~\ref{sec:valid}).
Nevertheless, the true pion distribution amplitude may well be a
``hybrid'' of the type
\hbox{
$
    \Phi _{\pi}^{\rm true}
 =
    90\% \Phi _{\pi}^{\rm as}
  + 9\% \Phi _{\pi}^{\rm CZ}
  + 1\% C_{4}^{(3/2)}
$},
where the mixing ensures a broader shape with the fourth-order,
``Mexican hat''-like, Gegenbauer polynomial $C_{4}^{(3/2)}$, being added
in order to cancel the dip of $\Phi _{\rm CZ}$ at $x=1/2$.

First tasks from instanton-based approaches show that the extracted
pion distribution amplitudes are very close to, albeit somewhat broader
than, the asymptotic form \cite{Dor96,ADT99,PPRWG98}.
Similar results were also obtained using nonlocal condensates
\cite{MR86,BM95}.
The discussion of non-asymptotic pion distribution amplitudes will be
conducted elsewhere.

\section{P\lowercase{ion form factor to order}
$\left(\alpha _{\lowercase{\rm s}}^{\lowercase{\rm an}}
               (Q^{2})\right)^{2}$}
\label{sec:piffNLO}

Next, we generalize our calculation of the hard contribution to the pion
form factor by taking into account the perturbative correction to
$T_{\rm H}$ of order $\alpha _{\rm s}^{2}$, using the results obtained
in \cite{FGOC81,DR81,BT87,MNP98}, in combination with our analytical,
i.e., IR-finite (IRF) factorization and renormalization scheme.

To be precise, we only include the NLO corrections to $T_{\rm H}$,
leaving NLO corrections to the evolution of the pion distribution
amplitude aside.
The reason is that for the asymptotic distribution amplitude, at issue
here, these corrections are tiny, appearing first at NLO
\cite{Mue94,MNP98}.
For sub-asymptotic distribution amplitudes, however, evolutional
corrections \cite{Mue94} have to be taken into account.
Strictly speaking, the calculation below is incomplete, the reason
being that the transverse degrees of freedom in the NLO terms of
$T_{\rm H}$ have been neglected, albeit the intrinsic ones in the
wave functions have been taken into account -- in contrast to other
approaches \cite{MNP98}.
Hence, our prediction should be regarded rather as an {\it upper limit}
for the size of the hard contribution to the pion form factor than as
an exact result.
Taking into account the $k_{\perp}$-dependence of $T_{\rm H}$ at NLO,
as we did for the leading part, this result might be somewhat reduced
as shown for the pion in \cite{JK93} and for the nucleon in
\cite{BJKBS95} (for a comprehensive discussion of $k_{\perp}$ effects,
we refer to \cite{Ste99}), though we expect that due to IR-finiteness,
this reduction should be rather small and the quality of our predictions
almost unchanged.
Note in this context that we always refer to the asymptotic distribution
amplitude of the pion.
Broadening the pion distribution amplitude would lead to a larger
normalization of (form-factor) magnitudes.
We would also like to emphasize that other higher-twist contributions
of non-perturbative origin, as those mentioned before, may also raise
the magnitude of the form factor.
However, such contributions are not on the focus of the present work.

Applying these assumptions, Eq.~(\ref{eq:pifofafin}) extends to NLO
to read
\begin{eqnarray}
\everymath{displaystyle}
  F_{\pi}\left( Q^{2} \right)
& = &
  16 A^{2} \pi C_{\rm F}
  \left(
        \frac{f_{\pi}/\sqrt{2}}{2\sqrt{N_{\rm c}}}
  \right)^{2}
  \int_{0}^{1} dx \int_{0}^{1} dy
  \int_{0}^{\infty} b \,db \,
  \alpha_{\rm s}^{\rm an}\left( \mu_{\rm R}^{2}\right)
  \Phi _{\rm as}(x) \Phi _{\rm as}(y)
\nonumber \\
&& \times
  \exp\left[
            - \frac{b^{2}\left( x\bar x + y\bar y \right)}
              {4\beta _{\rm as}^{2}}
      \right]
  \exp{ \left[-\beta _{\rm as}^{2} m_{\rm q}^{2}
  \left(\frac{1}{x\bar x} + \frac{1}{y\bar y}\right)
        \right]}
\nonumber \\
&& \times
  K\left(\sqrt{x y}Qb\right)
  \exp\left( - S\left(x, y, b, Q, C_{1}, C_{2}, C_{4} \right)
      \right)
\nonumber \\
&& \times
  \left[
        1 + \frac{\alpha _{\rm s}^{\rm an}}{\pi}
            \left( f_{\rm UV}\left( x, y, Q^{2}/\mu _{\rm R}^{2} \right)
          + f_{\rm IR}\left(x, y, Q^{2}/\mu _{\rm F}^{2}\right)
          + f_{\rm C}(x, y)
            \right)
  \right] \; ,
\label{eq:NLOpiff}
\end{eqnarray}
%Eq (42) Pion form factor with NLO terms in T_H
where the Sudakov form factor, including evolution, is given by
Eq.~(\ref{eq:fullSudakov}), $\mu _{\rm F}= C_{1}/b$, and the functions
$f_{i}$ are taken from \cite{MNP98}.
They are given by
\begin{eqnarray}
\nonumber
  f_{\rm UV}\left(x, y, Q^{2}/\mu _{\rm R}^{2}\right)
& = &
  \frac{\beta _{0}}{4}
    \left(\frac{5}{3}
  - \ln\left(\bar{x}\bar{y}\right)
  + \ln\frac{\mu _{\rm R}^{2}}{Q^{2}}\right) \; , \\
\nonumber
  f_{\rm IR}\left(x, y, Q^{2}/\mu _{\rm F}^{2}\right)
& = &
  \frac{2}{3}
  \left(
        3 + \ln\left(\bar{x}\bar{y}\right)
  \right)
         \left(
                 \frac{1}{2}\ln\left(\bar{x}\bar{y}\right)
               - \ln\frac{\mu _{\rm F}^{2}}{Q^{2}}
         \right) \; , \\
\nonumber
  f_{\rm C}(x,y)
& = &
  \frac{1}{12}
  \left[- 34 + 12 \ln\left(\bar{x}\bar{y}\right)
        + \ln x \ln y \right. \\
\nonumber
&& + \ln\bar{x}\ln\bar{y} - \ln x\ln\bar{y} - \ln\bar{x}\ln y \\
&& \left. + (1-x-y)H(x,y) + R(x,y)\right]
\label{eq:fis}
\end{eqnarray}
%Eq (43) Definition of functions f_i entering the NLO calculation
and are related to UV and IR poles, as indicated by corresponding
subscripts, that have been removed by dimensional regularization along
with the associated constants
$
 \ln (4\pi ) - \gamma _{\rm E}
$, whereas $f_{\rm C}(x,y)$ is scale-independent.
In evaluating expression $f_{\rm C}$ in (\ref{eq:fis}), we found it
particularly convenient to use the representation of the function
$H(x,y)$ given by Braaten and Tse \cite{BT87},
\begin{equation}
  H(x,y)
=
  \frac{1}{1-x-y}
  \left[
          {\rm Li}_{2}(\bar x)
        + {\rm Li}_{2}(\bar y)
        - {\rm Li}_{2}(x)
        - {\rm Li}_{2}(y)
        + \ln x \ln y
        - \ln \bar x \ln \bar y
  \right] \; ,
\label{eq:H}
\end{equation}
%Eq (44) Braaten and Tse's expression for function H(x,y)
where again ${\rm Li}_{2}$ denotes the dilogarithm function.
For the function $R(x,y)$ we have used the expression derived by Field
{\it et al.} \cite{FGOC81}, except at point $x\approx y$, where we
employed the Taylor expansion displayed below:
\begin{eqnarray}
\everymath{dispalystyle}
  R(x,y)
& = &
  \frac{1}{3\,\left( -1 + y \right)\, y^{2}}
  \Bigl[
        \left( - 1 + 33\,y - 45\, y^{2} + 13\,y^{3} \right) \,
         \ln \bar y
\nonumber \\
&& \phantom{\frac{1}{3\,\left(-1+y\right)\,y^{2}}}
               + y\,\left( -1 + y + ( 9 - 13\,y )\,y\,\ln y
        \right)
   \Bigr]
\nonumber \\
&&
 + \frac{x-y}{ 3\,(-1+y)^{2}\,y^{3} }
   \Bigl[
         ( - 1 + y )^{2}\, ( -1 + 16\,y )\, \ln \bar y
\nonumber \\
&& \phantom{\frac{x-y}{ 3\,(-1+y)^{2}\,y^{3}}+}
           + y\, \left( -1 + 13\,y - 12\,y^{2} + 2\,y^{2}\,\ln y
                 \right)
   \Bigl]
\nonumber \\
&&
 + \frac{ ( x - y )^{2} }{ 30\, ( -1 + y )^{3}\,y^{4} }
    \Bigl[
           ( - 1 + y )^{3}\, \left( 9 - 148\,y + 9\,y^{2}\right) \,
           \ln \bar y
\nonumber \\
&& \phantom{\frac{(x-y)^{2}}{30\,(-1+y)^{3}\,y^{4}}+}
          - y \left( 9 - 148\,y + 328\,y^{2} - 189\,y^{3}
          + y^{3}\,( 5 + 9\,y )\,\ln y \right)
    \Bigr] \; .
\label{eq:TaylorR}
\end{eqnarray}
%Eq (45) Taylor expansion around x=y of R(x,y)
Note that this expression does not reproduce its counterpart in
\cite{FGOC81}.
It must be remarked once again that evaluating Eq.~(\ref{eq:NLOpiff})
there is an analytization ambiguity similar to that encountered in
the calculation of the Sudakov exponent.
This question will be addressed elsewhere.

Having developed in detail the theoretical apparatus, let us now turn to
the concrete (numerical) calculation of the pion form factor at NLO.

\section{N\lowercase{umerical analysis}}
\label{sec:valid}

This section implements factorization on the numerical level, thus
providing the bridge between the analytic framework, developed and
discussed in the previous sections, and numerical calculations to
follow in the subsequent section.
This is done by appropriately defining the accessible phase space
regions (kinematic integrals) of transverse momenta (or equivalently
transverse distances $b$) for gluon emission in each regime, making
explicit the inherent kinematical restrictions on the momenta of
hard (soft) gluons due to factorization.
The numerical analysis below updates and generalizes our previous
investigations in Refs. \cite{SSK98,SSK98a}.

In order to set up a reliable algorithm for the numerical evaluation of
the expressions presented above, we have to ensure that this is done in
kinematic regions where use of fixed-order or re-summed perturbation
theory is legal.
Further, expedient restrictions have to be imposed to avoid double
counting of gluon corrections by carefully defining the validity domain
of each contribution to the pion form factor, in correspondence with
Fig.~\ref{fig:qcdreg}.
These kinematic constraints are compiled below.

\begin{center}
{\small{\it  Kinematic cuts}}
\end{center}
\begin{enumerate}
\item $C_{1}/b > \Lambda _{\rm QCD}$; otherwise the whole Sudakov
      exponent $\exp (-S)$ (cf.~Eq.~(\ref{eq:fullSudakov})) is continued
      to zero because this large-$b$ region is properly taken into
      account in the wave functions.
      This condition excludes from the re-summed perturbation theory soft
      gluons with wavelengths larger than $C_{1}/\Lambda$, which should
      be treated non-perturbatively.
      In other words, it ensures the separation (factorization) of the
      effective sub-sector from the genuine non-perturbative regime
      (cf.~Fig.~\ref{fig:qcdreg}).
\item $C_{2}\xi Q > C_{1}/b$; otherwise each Sudakov exponent
      $\exp \left[- s\left(\xi,b,Q,C_{1},C_{2}\right)\right]$ in
      Eq.~(\ref{eq:fullSudakov}) is ``frozen'' to unity because this
      small-$b$ region is dominated by low orders of perturbation theory
      rather than by the re-summed perturbation series, and consequently
      contributions in this region should be ascribed to higher-order
      corrections to $T_{\rm H}$, which we have taken into account
      explicitly at NLO.
      This condition establishes proper factorization between re-summed
      and fixed-order perturbation theory and helps avoid double
      counting of such contributions (always working in the gauge
      $A^{+}=0$).
      Yet evolution is taken into account to match the scales in our
      ``gliding'' factorization scheme.
\item $C_{4}f(x,y)Q > C_{1}/b$; otherwise the evolution time
      $\tau \left( C_{1}/b, \mu _{\rm R} \right)$ in
      Eq.~(\ref{eq:fullSudakov}) is contracted to zero, i.e., evolution
      is ``frozen''.
      The renormalization scale should be at least equal to the
      factorization scale, so that the running coupling has always
      arguments in the range controlled by (re-summed or fixed-order)
      perturbation theory.
\item $C_{4}f(x,y)Q > C_{2}\xi Q$; otherwise evolution to that scale
      is ``frozen'' because this region is appropriately accounted for
      by the Sudakov contribution.
      This helps avoiding double counting of terms which belong to the
      re-summed rather than to the fixed-order perturbation theory.
      (No overlap at the boundary characterized by the scale $\mu ^{2}$
      in Fig.~\ref{fig:qcdreg}).
\item $C_{4}f(x,y)Q > C_{1}/b$; otherwise the two scales
      $\mu _{\rm R}=\mu _{\rm F}=C_{1}/b$
      are identified in the function $f_{\rm UV}(x,y)$
      (by the same reasoning as above).
      If $\mu _{\rm R}\leq \Lambda _{\rm QCD}$, then
      $f_{\rm UV}(x,y)$ is set equal to zero.
\item $C_{1}/b > \Lambda _{\rm QCD}$; otherwise the function
      $f_{\rm IR}(x,y)$ is set equal to zero.
      The last two restrictions exclude contributions from perturbative
      terms when they are evaluated in the non-perturbative kinematic
      domain.
\end{enumerate}

\begin{figure}
\tighten
\[
\psfig{figure=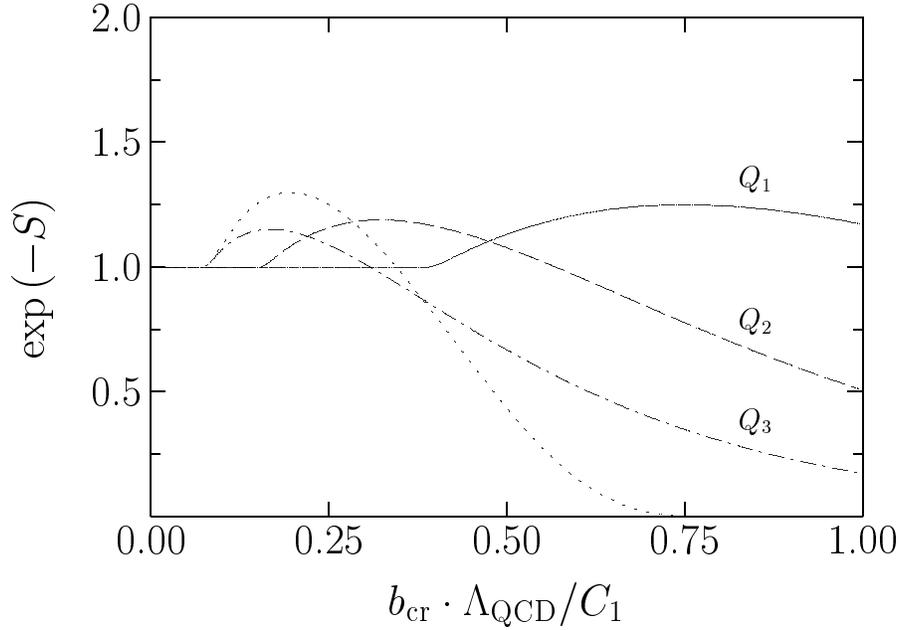,%
      bbllx=148pt,bblly=480pt,bburx=390pt,bbury=690pt,%
      height=7.0cm,silent=}
\]
\vspace{1.5cm}
\caption[fig:sudall]
        {\tenrm Behavior of the Sudakov form factor with respect to
         the transverse separation $b$ for three representative values
         of the momentum transfer $Q^{2}$:
         $Q_{1}=2$~GeV, $Q_{2}=5$GeV, and $Q_{3}=10$~GeV, with all
         $\xi _{i}=1/2$, and where we have set
         $C_{1}=2{\rm e}^{-\gamma _{\rm E}}$,
         $C_{2}={\rm e}^{-1/2}$ and
         $\Lambda _{\rm QCD}=0.242~{\rm GeV}$.
         The dotted curve shows the result obtained with
         $\alpha _{\rm s}^{\overline{\rm MS}}$, and
         $\Lambda _{\rm QCD}=0.2~{\rm GeV}$ for $Q_{2}=5$~GeV, using
         the same set of $C_{i}$ as before.
         Notice that in this case, evolution is limited by the
         (renormalization) scale
         $\mu_{\rm R}=t=\{ {\rm max}\sqrt{xy}\, Q, C_{1}/b \}$,
         as proposed in \cite{LS92}.
         However, the enhancement at small $b$-values due to the quark
         anomalous dimension is not neglected here.
\label{fig:sudalepj}}
\end{figure}

To illustrate the difference in technology between approaches
employing the conventional expression for the full Sudakov exponent
\cite{BS89,LS92}, on one hand, and our analysis, on the other, we
show $\exp (-S)$ graphically in Fig.~\ref{fig:sudalepj} for three
different values of the momentum transfer and $\xi = 1/2$.
In contrast to Li and Sterman \cite{LS92}, the evolutional contribution
is not cut-off at unity, whenever $C_{2}\xi Q < C_{1}/b$.
The dotted curve shows the result for Eq.~(\ref{eq:oldsudakov}) without
this cutoff.
One infers from this figure that their suggestion to ignore the
enhancement due to the anomalous dimension does not apply in our
case because the IR-modified Sudakov form factor is not so rapidly
decreasing as $b$ increases, owing to the IR-finiteness of
$\alpha _{\rm s}^{\rm an}$.
Indeed, as $Q$ becomes smaller, $\exp (-S)$ remains constant and fixed
to unity for increasing $b$, providing enhancement only in the
large-$b$ region before it reaches the kinematic boundary
$C_{1}/b=\Lambda _{\rm QCD}$, where it is set equal to zero.
As a result, for small $Q$-values, like $Q_{1}=2$~GeV, the enhancement
due to the quark anomalous dimension cannot be associated with
higher-order corrections to $T_{\rm H}$, since it operates at larger
$b$-values, and for that reason it should be taken into account in the
Sudakov contribution.
Only for asymptotically large $Q$ values, when the IR-modified Sudakov
form factor and the conventional one become indistinguishable, the
evolutional enhancement becomes a small effect -- strictly confined in
the small-$b$ region -- and can be safely ignored.
On the other hand, because the Sudakov exponent is bounded at fixed
$Q^{2}$, the Sudakov exponential remains finite until the edge of
phase space, $C_{1}/b_{\rm cr} \lesssim \Lambda _{\rm QCD}$, also
providing IR enhancement.
This behavior is best appreciated by comparing the dashed and dotted
curves, both at $Q_{2}=5$~GeV, in Fig.~\ref{fig:sudalepj}.

%%%%%%%%%%%%%%%%%%%%%%%%%%%%%%%%%%%%%%%%%%%%%%%%%%%%%%%%%%%%%%%%%%%%%
%                            F I G U R E  3                         %
%                         \label{fig:sudalepj}                      %
%%%%%%%%%%%%%%%%%%%%%%%%%%%%%%%%%%%%%%%%%%%%%%%%%%%%%%%%%%%%%%%%%%%%%
%
%%%%%%%%%%%%%%%%%%%%%%%%%%%%%%%%%%%%%%%%%%%%%%%%%%%%%%%%%%%%%%%%%%%%%

The behavior of the Sudakov form factor, we stress, shows that
power-induced sub-leading logarithmic corrections are relevant in the
range of currently probed momentum-transfer values.
Hence, the advantage of employing such a scheme to calculate hadronic
observables, for instance the pion form factor, is that the hard
(perturbative) contribution is enhanced, relative to the calculation
in the $\overline{\rm MS}$ scheme, and the self-consistency of the
perturbative treatment towards lower $Q^{2}$ values, where it is
not justified, is significantly improved (no enhancement caused by the
Landau obstruction; hence better scaling).

This is because the range in which soft gluons build up the Sudakov form
factor is enlarged and inhibition of bremsstrahlung sets in at larger
$Q^{2}$.
Let us mention in this context that power contributions in the radiative
corrections to the meson wave function could lead to suppression of soft
gluon emission at large transverse distance $b$.
Indeed, Akhoury, Sincovics, and Sotiropoulos \cite{ASS98} have re-summed
such power corrections, associated to IR renormalons, with the aid of an
effective gluon mass.
They found Sudakov-type suppression on top of the Sudakov suppression
discussed so far.
The discussion of such IR-renormalon-based contributions in conjunction
with our IR-finite approach will be presented elsewhere.

\section{P\lowercase{henomenology}}
\label{sec:pheno}

\begin{figure}
\tighten
\centerline{\epsfig{file=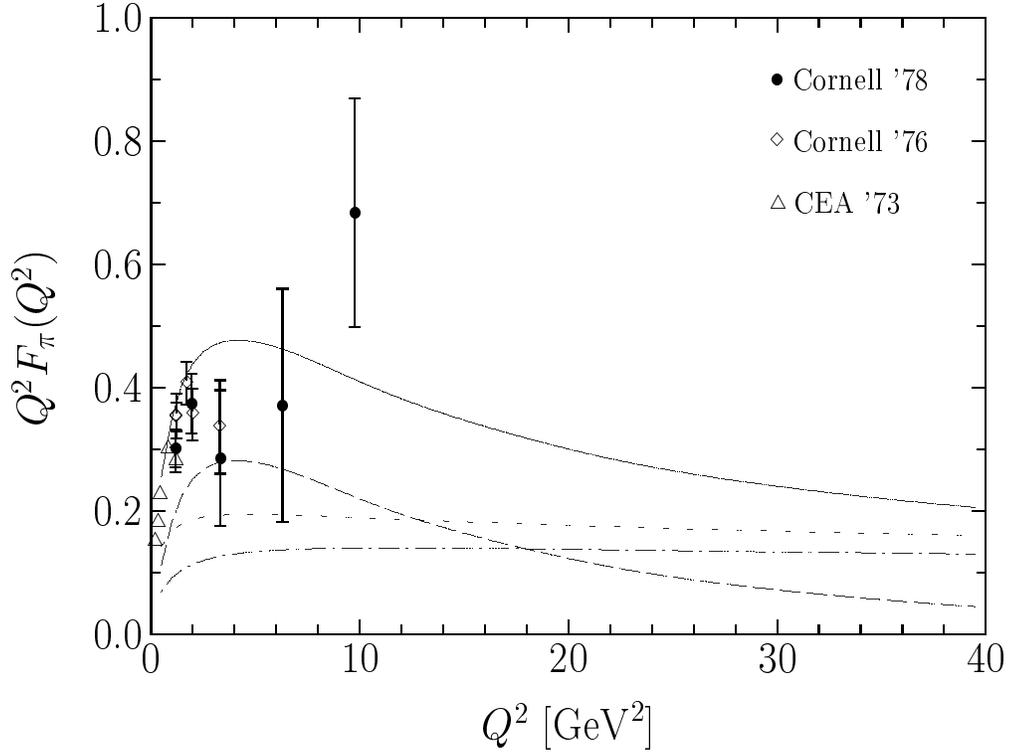,height=10cm,width=14cm,silent=}}
\vspace{1cm}
\caption[fig:pionffBLM]
        {\tenrm Space-like pion form factor calculated within our
         IR-finite scheme with $\Phi _{\rm as}$ and including an
         effective (constituent-like) quark mass of
         $m_{\rm q}=0.33$~GeV in the pion wave function.
         The broken lines show the IR-enhanced hard contributions
         obtained with our scheme using commensurate scale setting:
         LO calculation (dashed-dotted line); NLO calculation
         (dotted line).
         The dashed line gives the result for the soft, Feynman-type
         contribution, computed with $m_{\rm q}=0.33$~GeV in the pion
         wave function.
         The solid line represents the sum of the NLO hard contribution
         and the soft one.
         The data are taken from \cite{Bro73,Beb76}.
\label{fig:piffepj}}
\end{figure}

Let us now present phenomenological applications of our scheme.
Using the techniques discussed above, we obtain for the electromagnetic
pion form factor the theoretical predictions shown in
Fig.~\ref{fig:piffepj}.
A set of constants \hbox{$C_{i}$, $(i=1,2,3)$} which eliminate
artifacts of dimensional regularization, while practically preserving
the matching between the re-summed and the fixed-order calculation, are
given in Table \ref{tab:setsC_i} in comparison with other common choices
of these constants.
Moreover, this factorization scale setting enables us to naturally link
our scheme to the BLM commensurate scale method \cite{BLM83} in
fixing the renormalization scale.
Indeed, since the adopted value of $C_{1}$ eliminates both the log
term in the $K$-factor (see Eq.~(\ref{eq:Kfactor})), which contains the
$\beta$-function, and also the scheme-dependent term $B$ in the cusp
anomalous dimension (see Eq.~(\ref{eq:funBnlo})), this choice
corresponds to a conformally invariant framework with $\beta _{0}=0$,
and therefore connects to the commensurate scale procedure.
Hence, we set $C_{4}=C_{2}\exp{(-5/6)}$, which, for our choice of
$C_{2}=\exp{(-1/2)}$, rescales $Q$ in the $\overline{\rm MS}$
scheme, we use, by a factor of $\exp{(-4/3)}$.
In addition, to avoid large kinematical corrections due to soft gluon
emission, we set $f(x,y)=\sqrt{xy}$ to link the renormalization
scale to the typical momentum flow in the gluon propagators \cite{KT82}.
In this way, scheme and renormalization scale ambiguities are
considerably reduced, as the theoretical predictions are evaluated at a
physical momentum scale:
\begin{equation}
  \mu _{\rm BLM}
=
  \mu _{\rm R} \exp (-5/6) \; ,
\label{eq:renscaleBLM}
\end{equation}
%Eq (46) Commensurate renormalization scale a la BLM
where
\begin{equation}
  \mu _{\rm R}
=
  C_{4} f(x,y) Q
=
  C_{4}\sqrt{xy} Q \; .
\label{eq:IRFrenscale}
\end{equation}
%Eq (47) IRF renormalization scale setting
We emphasize, however, that these favored values of the scheme constants
by no means restrict the validity of our numerical analysis.
They merely indicate the anticipated appropriate choice of the
factorization and renormalization scales with respect to observables and
theoretical self-consistency.
Other choices of these parameters do not change the qualitative features
of our predictions.

%%%%%%%%%%%%%%%%%%%%%%%%%%%%%%%%%%%%%%%%%%%%%%%%%%%%%%%%%%%%%%%%%%%%%
%                          T A B L E  2                             %
%                      \label{tab:setsC_i}}                         %
%%%%%%%%%%%%%%%%%%%%%%%%%%%%%%%%%%%%%%%%%%%%%%%%%%%%%%%%%%%%%%%%%%%%%
\begin{table}
\caption[tab:coefc_i]
        {Different sets of coefficients $C_{i}$ and values of the
         $K$-factor and the quantity (cf.~Eq.~(\ref{eq:funBnlo}))
         $\kappa = \ln \left(
          C_{1}^{2}\;{\rm e}^{2\gamma_{E}-1} /4\, C_{2}^{2}
                       \right)$,
         corresponding to different factorization and renormalization
         prescriptions. The choice of $C_{4}$ in this work corresponds
         to BLM-type commensurate scale fixing.
\label{tab:setsC_i}}
\begin{tabular}{lcccccc}
            & \multicolumn{4}{c}{Scheme parameters $C_{i}$} \\
 Choice & $C_{1}$ &
 $C_{2}=\frac{1}{\sqrt{2}}C_{2}^{\rm CSS}$ \cite{CS81} &
 $C_{3}$ & $C_{4}$ & $K$ & $\kappa$ \\
\hline
 canonical & $2\exp \left(-\gamma _{\rm E}\right)$ & $1/\sqrt{2}$ &
 $2\exp \left(-\gamma _{\rm E}\right)$ & -- & 4.565 & -0.307 \\
 SSK \cite{SSK98} &
 $\exp\left[-\frac{1}{2}\left(2\gamma _{\rm E}-1\right)\right]$ &
 $1/\sqrt{2}$ &
 $\exp\left[-\frac{1}{2}\left(2\gamma _{\rm E}-1\right)\right]$ & -- &
 2.827 & 0 \\
 this work  & $2\exp \left(-\gamma _{\rm E}\right)$ &
 $\exp \left(-1/2\right)$ & $2\exp \left(-\gamma _{\rm E}\right)$ &
 $\exp \left(-4/3\right)$ & 4.565 & 0 \\
\end{tabular}
\end{table}
%T A B L E  2
%
%%%%%%%%%%%%%%%%%%%%%%%%%%%%%%%%%%%%%%%%%%%%%%%%%%%%%%%%%%%%%%%%%%%%%

%%%%%%%%%%%%%%%%%%%%%%%%%%%%%%%%%%%%%%%%%%%%%%%%%%%%%%%%%%%%%%%%%%%%%
%                            F I G U R E  4                         %
%                         \label{fig:piffepj}                       %
%%%%%%%%%%%%%%%%%%%%%%%%%%%%%%%%%%%%%%%%%%%%%%%%%%%%%%%%%%%%%%%%%%%%%
%
%%%%%%%%%%%%%%%%%%%%%%%%%%%%%%%%%%%%%%%%%%%%%%%%%%%%%%%%%%%%%%%%%%%%%

\begin{figure}
\tighten
\centerline{\epsfig{file=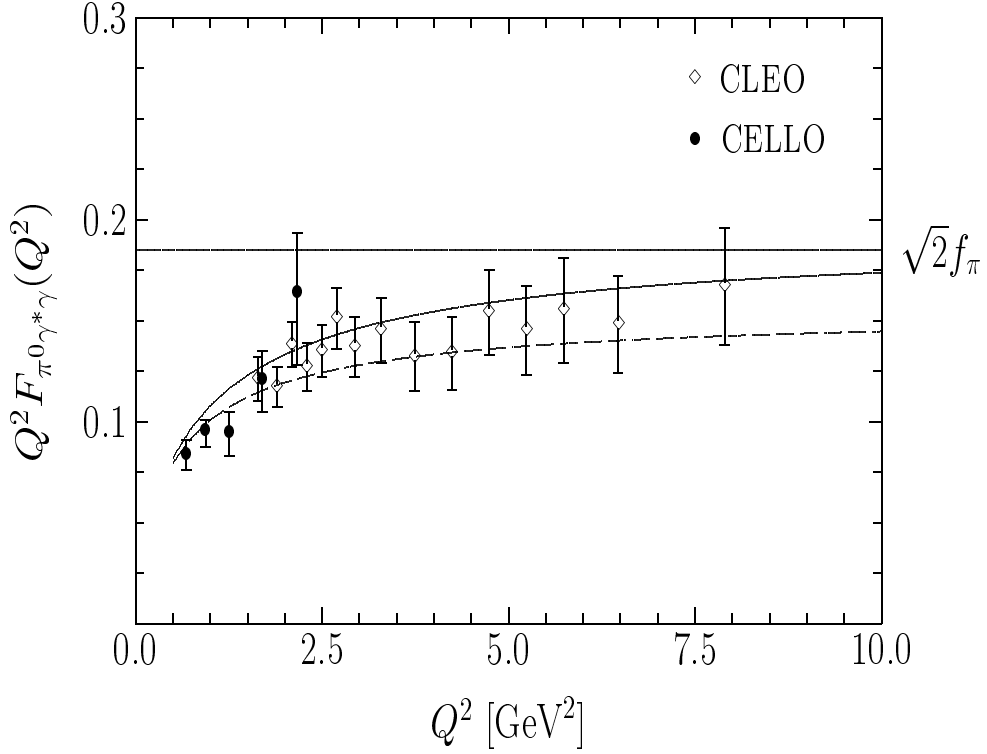,height=10cm,width=14cm,silent=}}
\vspace{1cm}
\caption[fig:piongamma]
        {\tenrm Pion-photon transition form factor for the asymptotic
         distribution amplitude, calculated in the IRF scheme.
         The dashed line shows the prediction for the BHL ansatz
         for the pion wave function which includes an effective
         (constituent-like) quark mass of $m_{\rm q}=0.33$~GeV.
         Commensurate scale setting is used, i.e.,
         $C_{4}=C_{2}\exp{(-5/6)}$ with $C_{2}=\exp{(-1/2)}$.
         For comparison, we also show the prediction (solid line)
         obtained without a quark mass and using a non-commensurate
         renormalization scale ($C_{4}=C_{2}=\exp{(-1/2)}$).
         The horizontal line represents the asymptotic behavior.
         The data are taken from \cite{CLEO98,CELLO91}.
\label{fig:pigaepj}}
\end{figure}

Before we proceed with the discussion of these results, let us first
present the theoretical prediction for the pion-photon transition form
factor $F_{\pi\gamma ^{*}\gamma}(Q^{2},q^{2}=0)$ in which one of the
photons is highly off-shell and the other one is close to its
mass-shell.
In leading perturbative order, this form factor is given by the
expression (cf.~\cite{JKR96})
\begin{eqnarray}
\everymath{displaystyle}
  F_{\pi\gamma}\left(Q^{2}\right)
& = &
   \frac{A}{\sqrt{3}\pi}
   \int_{0}^{1} dx \int_{0}^{\infty} db \,
   b \, \frac{(f_{\pi}/\sqrt{2}) \Phi _{\rm as}(x)}{2\sqrt{N_{\rm c}}}
   \exp\left(-x\bar{x}b^{2}/4\beta _{\rm as}^{2}
       \right)
\nonumber \\
&& \times
   \exp{
        \left(-\beta _{\rm as}^{2} m_{\rm q}^{2} \frac{1}{x\bar x}
        \right)
        }
   \left(4\pi K_{0}\left(\sqrt{\bar{x}}\,bQ\right)\right)
   e^{-S_{\pi\gamma}} \; ,
\label{eq:ffpigamma}
\end{eqnarray}
%Eq (48) Pion-photon transition form factor
where the Sudakov exponent, including evolution, has the form
\begin{equation}
  S_{\pi\gamma}\left( x, \bar{x}, b, Q, C_{1}, C_{2}, C_{4} \right)
=
    s(x,b,Q) + s(\bar{x},b,Q)
  - 4 \tau \left( \frac{C_{1}}{b}, \mu _{\rm R} \right) \; .
\label{eq:sudpigamma}
\end{equation}
%Eq (49) Full Sudakov exponent for pion-photon transition form factor
The main difference relative to the previous case is that this form
factor contains only one pion wave function, whereas the associated
hard-scattering part, being purely electromagnetic at this order, does
not depend directly on $\alpha _{\rm s}$.
The only dependence on the (running) strong coupling enters through the
anomalous dimensions in the Sudakov form factor.
The result of this calculation is displayed in Fig.~\ref{fig:pigaepj}.

%%%%%%%%%%%%%%%%%%%%%%%%%%%%%%%%%%%%%%%%%%%%%%%%%%%%%%%%%%%%%%%%%%%%%
%                           F I G U R E  5                          %
%                        \label{fig:pigaepj}                        %
%%%%%%%%%%%%%%%%%%%%%%%%%%%%%%%%%%%%%%%%%%%%%%%%%%%%%%%%%%%%%%%%%%%%%
%
%%%%%%%%%%%%%%%%%%%%%%%%%%%%%%%%%%%%%%%%%%%%%%%%%%%%%%%%%%%%%%%%%%%%%

All constraints on kinematics set forward in the numerical evaluation
of the electromagnetic pion form factor are relevant to this case too,
except the requirement which deals specifically with the choice of the
renormalization scale, which now is set equal to
$\mu _{\rm R} = C_{4} x Q$
because only one pion wave function is involved.
Another reasonable choice would be
$\mu _{\rm R} = C_{4} \sqrt{x\bar x} Q$,
which entails evolution to a lower scale, hence reducing evolutional
enhancement through
$\tau \left(C_{1}/b,\mu _{\rm R} \right)$
by approximately $6\%$.

Let us now discuss these effects more systematically.

It is obvious from Fig.~\ref{fig:piffepj} that the IR-enhanced hard
contribution to $F_{\pi}(Q^{2})$ with optimized choice of scales is
providing a sizeable fraction of the magnitude of the form factor --
especially at NLO.
This behavior is IR stable from low to high $Q^{2}$ values, exhibiting
almost exact scaling (in accordance with the nominal scaling of the
leading twist prediction), which shows that the analytic coupling is
almost constant in a wide range of $Q^{2}$ values.
In contrast to other approaches \cite{TL98,MNP98,BKM99}, which involve
a running $\alpha _{\rm s}$ coupling without an IR-fixed point,
there is no artificial rising at low $Q^{2}$ of the of the hard form
factor, resulting from the unphysical Landau pole.
Furthermore, by employing a commensurate scale setting to fix the
renormalization point, the scheme and renormalization-prescription
dependence of our predictions has been minimized.
In addition, the imposed kinematical constraints in our numerical
analysis ensure that the contributions, originating from different
phase space regions, do not overlap to give rise to double counting.

\begin{figure}
\tighten
\centerline{\epsfig{file=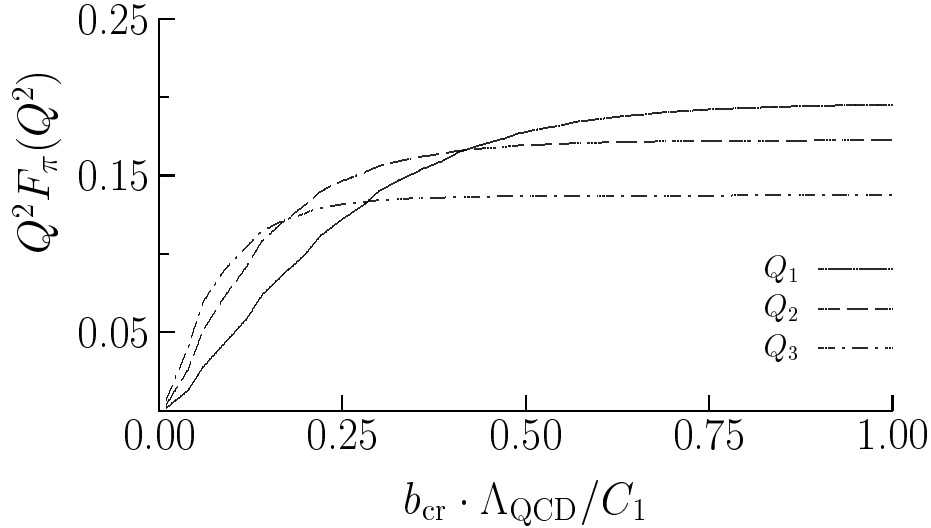,height=7cm,width=13cm,silent=}}
\vspace{1cm}
\caption[fig:pionsaturation]
        {\tenrm Saturation behavior of the pion electromagnetic
         form factor, calculated in the IRF scheme at NLO with
         commensurate scale setting and including a mass term
         (with $m_{\rm q}=0.33$~GeV) in the BHL ansatz for the
         soft pion wave function. The scheme parameters are defined
         in Table \ref{tab:setsC_i}. Here $b_{\rm cr}$ denotes the
         integration cutoff over transverse distances in
         Eq.~\ref{eq:NLOpiff}. The momentum transfer values are as
         in Fig.~\ref{fig:sudalepj}.
\label{fig:satcomas}}
\end{figure}

The reduced sensitivity of the perturbatively calculated hard form
factor to the endpoint region, is also reflected in its saturation
behavior.
One sees from Fig.~\ref{fig:satcomas} that the bulk of the scaled form
factor
$Q^{2}F_{\pi}\left(Q^{2}\right)$
is already accumulated below
$b_{\rm cr}\Lambda _{\rm QCD}/C_{1}\leq 0.5$,
i.e., for short transverse distances, where the application of
perturbative QCD is self-consistent.
All curves shown rise very steeply to their full height at the
integration cutoff
$b_{\rm cr}=C_{1}/\Lambda _{\rm QCD}$,
beyond which they flatten out, indicating that remaining contributions
are truly of nonperturbative origin.
One observes that the perturbative treatment starts to be reliable
already at $Q^{2}=4$~GeV${}^{2}$ and improves further, albeit not
dramatically, as the momentum transfer increases.
A fast saturation behavior in the small $b$-region, where contamination
with nonperturbative contributions is still not serious and the coupling
constant is small, is considered as a standard to judge the
self-consistency of the perturbative method applied.
Though the Sudakov form factor contains considerable contributions
from gluons with wave lengths of the order of
$C_{1}/\Lambda _{\rm QCD}$ due to the IR finiteness of the running
coupling (see Fig.~\ref{fig:sudalepj}) -- especially at low momentum
transfer -- we realize that the form factor itself, i.e., the physical
observable, does not receive strong contributions from this endpoint
($b$) region.
Moreover, the form factor calculation does not receive large
contributions from the endpoint region in $x$ as well, as we use only
the asymptotic pion distribution amplitude (or such close to it).
Hence, from the theoretical point of view, the quality and
self-consistency of the perturbative treatment have been improved
relative to previous approaches \cite{LS92,JK93}.

%%%%%%%%%%%%%%%%%%%%%%%%%%%%%%%%%%%%%%%%%%%%%%%%%%%%%%%%%%%%%%%%%%%%%
%                           F I G U R E  6                          %
%                        \label{fig:satcomas}                       %
%%%%%%%%%%%%%%%%%%%%%%%%%%%%%%%%%%%%%%%%%%%%%%%%%%%%%%%%%%%%%%%%%%%%%
%
%%%%%%%%%%%%%%%%%%%%%%%%%%%%%%%%%%%%%%%%%%%%%%%%%%%%%%%%%%%%%%%%%%%%%

\begin{figure}
\tighten
\centerline{\epsfig{file=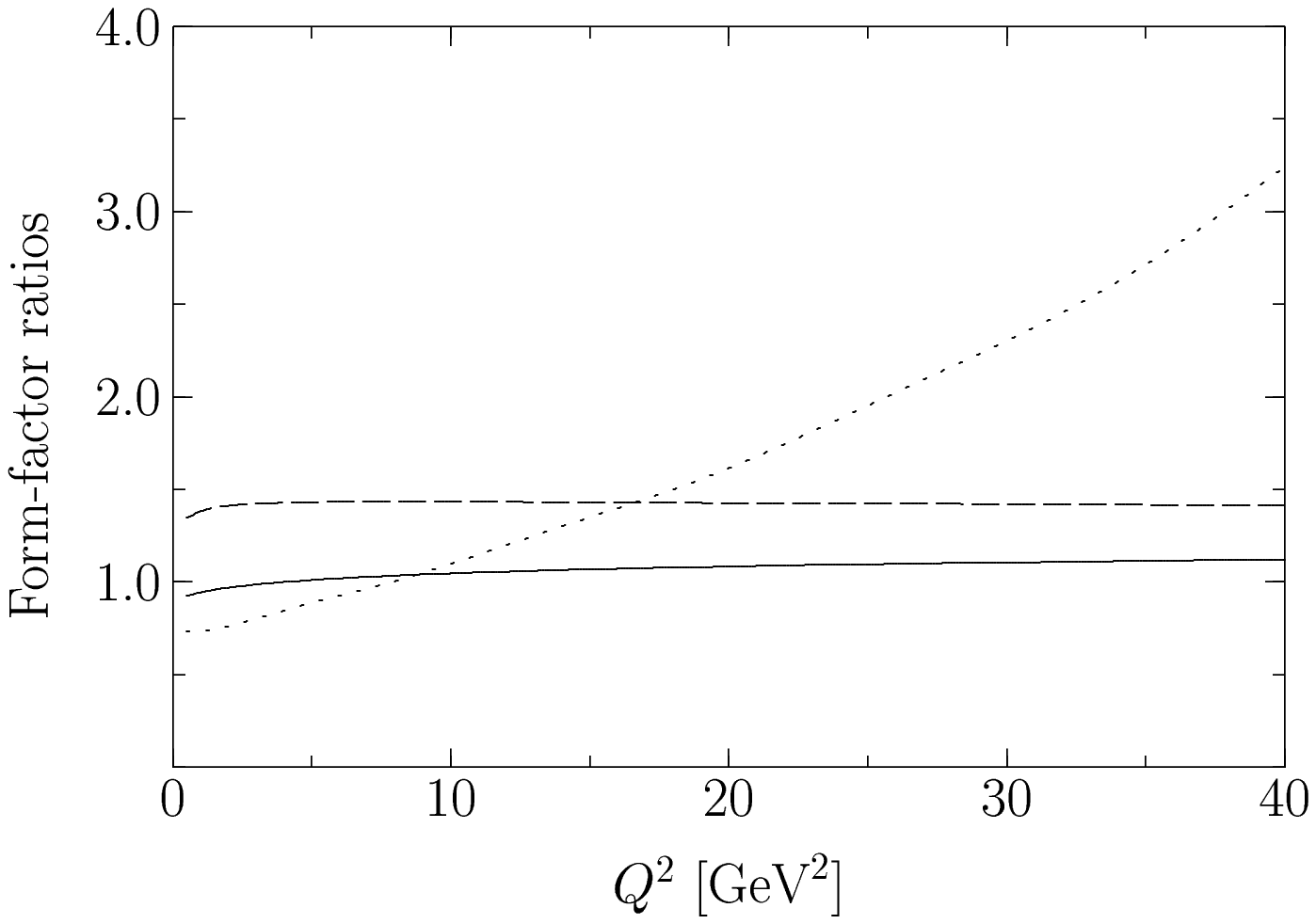,height=10cm,width=14cm,silent=}}
\vspace{1cm}
\caption[fig:div_effects]
        {\tenrm Effect of using a BHL-type ansatz for the pion wave
                function with an effective (constituent-like) quark
                mass of $m_{\rm q}=0.33$~GeV. The solid line plots
                the ratio
$
 {\left(F_{m_{\rm q}=0}^{\rm hard}/
 F_{m_{\rm q}\neq 0}^{\rm hard}\right)}\big|_{\rm comm}^{\rm NLO}
$
                and the dotted line the ratio
$
 {\left(F_{m_{\rm q}=0}^{\rm soft}/
 F_{m_{\rm q}\neq 0}^{\rm soft}\right)}\big|_{\rm comm}
$
                versus the momentum transfer $Q^{2}$.
                The dashed line shows the ratio
$
 {\left(F_{\rm comm}^{\rm hard}/
 F_{\rm non-comm}^{\rm hard}\right)}\big|_{m_{\rm q}\neq 0}^{\rm NLO}
$
                of the hard part of the pion form factor for a
                BLM commensurate scale setting relative to a
                conventional one with $C_{4}=C_{2}=\exp{(-1/2)}$
                (cf.~Table \ref{tab:setsC_i}).
\label{fig:effects}}
\end{figure}

Figure \ref{fig:effects} shows the influence of the effective quark mass
on the pion form factor.
The designations are as follows:
The solid line plots
$
 {\left(F_{m_{\rm q}=0}^{\rm hard}/
 F_{m_{\rm q}\neq 0}^{\rm hard}\right)}\big|_{\rm comm}^{\rm NLO}
$
and the dotted line
$
 {\left(F_{m_{\rm q}=0}^{\rm soft}/
 F_{m_{\rm q}\neq 0}^{\rm soft}\right)}\big|_{\rm comm}
$).
It is clearly obvious that the effect of the quark mass on the hard part
is negligible in size and does not depend on the variation in $Q^{2}$,
whereas the soft contribution gets significantly reduced as $Q^{2}$
grows.
The dashed line, standing for the expression
$
 {\left(F_{\rm comm}^{\rm hard}/
 F_{\rm non-comm}^{\rm hard}\right)}\big|_{m_{\rm q}\neq 0}^{\rm NLO}
$,
in the same figure quantifies the effect of using a commensurate scale
setting for the renormalization scale.
As one sees, this amounts to an enhancement factor of about 1.5.

%%%%%%%%%%%%%%%%%%%%%%%%%%%%%%%%%%%%%%%%%%%%%%%%%%%%%%%%%%%%%%%%%%%%%
%                           F I G U R E  7                          %
%                        \label{fig:effects}                        %
%%%%%%%%%%%%%%%%%%%%%%%%%%%%%%%%%%%%%%%%%%%%%%%%%%%%%%%%%%%%%%%%%%%%%
%
%%%%%%%%%%%%%%%%%%%%%%%%%%%%%%%%%%%%%%%%%%%%%%%%%%%%%%%%%%%%%%%%%%%%%

The advantages of our framework may become more transparent by
comparing our results with those obtained in other analyses.
This is done for the pion form factor in Table \ref{tab:piffval}.

%%%%%%%%%%%%%%%%%%%%%%%%%%%%%%%%%%%%%%%%%%%%%%%%%%%%%%%%%%%%%%%%%%%%%
%                             T A B L E  3                          %
%                         \label{tab:piffval}}                      %
%%%%%%%%%%%%%%%%%%%%%%%%%%%%%%%%%%%%%%%%%%%%%%%%%%%%%%%%%%%%%%%%%%%%%
\begin{table}
\caption[tab:valpiff]
        {Calculated pion form factor at two values of $Q^{2}$.
         The first two columns show the results obtained in the present
         work in comparison with those calculated by Jakob and Kroll
         (JK) \cite{JK93} (third column), and by Meli\'c, Ni\v zi\'c,
         and Passek (MNP) \cite{MNP98} (last two columns).
\label{tab:piffval}}
\begin{tabular}{cccccc}
         $Q^{2}$~[GeV${}^{2}$] & LO & LO+NLO & LO & LO & LO+NLO\\
         & (this work) & (this work) & (JK) & (MNP) & (MNP) \\
\hline
         4  & 0.128 & 0.191 & 0.08 & 0.131 & 0.211 \\
        10  & 0.137 & 0.186 & 0.08 & 0.109 & 0.164 \\
\end{tabular}
\end{table}
%T A B L E  3
%
%%%%%%%%%%%%%%%%%%%%%%%%%%%%%%%%%%%%%%%%%%%%%%%%%%%%%%%%%%%%%%%%%%%%%

Comparison of our values with those calculated by Jakob and Kroll
\cite{JK93} shows that the suppression of the hard part of the form
factor due to the inclusion of transverse degrees of freedom is
counteracted by the power-induced enhancement, amounting to an average
enhancement of about $50\%$ relative to their values.
This is achieved by using (almost) the same root mean square transverse
momentum of
${\langle {\kv}^{2}\rangle}^{1/2}=0.352$~GeV,
as in their analysis, and with a reasonable probability for the
valence Fock state of $P_{\rm{q\bar q}}=0.306$ (see Table
\ref{tab:parameters}).
The inclusion of an effective (constituent-like) quark mass in the
Gaussian ansatz for the distribution of intrinsic transverse momentum
in the pion wave function changes dramatically the fall-off behavior
of the soft contribution to the form factor, as compared to the JK
analysis, though its maximum size remains almost unchanged, and its
influence on the hard part is very small (cf.~Figs.~\ref{fig:piffepj}
and \ref{fig:effects}).
Indeed, one infers from Fig.~\ref{fig:piffepj} that
$F_{\pi}^{\rm soft}$ becomes equal to
$F_{\pi}^{\rm hard}$ already at
$Q^{2}\simeq 18$~GeV${}^{2}$ (LO result), or even at
$Q^{2}\simeq 12$~GeV${}^{2}$ when the NLO corrections are included.
This behavior of $F_{\pi}^{\rm soft}$ falls well in line with
the correct behavior of $\Psi _{\pi}^{\rm soft}$ for ${\kv}=0$
and $k_{3}\to-\infty$, restored by the mass term and the arguments on
the nonperturbative vacuum dynamics given above.

On the other hand, comparison with the values computed by Meli\'c,
Ni\v zi\'c, and Passek \cite{MNP98} at leading order, by completely
ignoring transverse degrees of freedom, reveals that in the $Q^{2}$
domain, where the influence of the Landau singularity has died out
(values for $Q^{2}=10$~GeV${}^{2}$ in Table \ref{tab:piffval}), there
is still enhancement of about $17\%$.
Comparing our results with theirs at next-to-leading order, we conclude
that our choice of scheme and renormalization scales is consistent with
a proper matching between gluon corrections, calculated on a
term-by-term perturbation expansion (NLO corrections to $T_{H}$), and
those due to the re-summed perturbative series (Sudakov form factor).
Therefore, double counting of such contributions in our scheme, if
any, must indeed be negligible.
Moreover, the scaling behavior of the calculated perturbative (hard)
form factor is considerably improved, complying with the nominal
scaling of the leading twist prediction.
Indeed, one observes (cf.~\ref{tab:piffval}) that the deviation from
exact scaling, associated with NLO evolutional corrections of the
asymptotic distribution amplitude, is, as stated before, negligible.

\begin{figure}
\tighten
\centerline{\epsfig{file=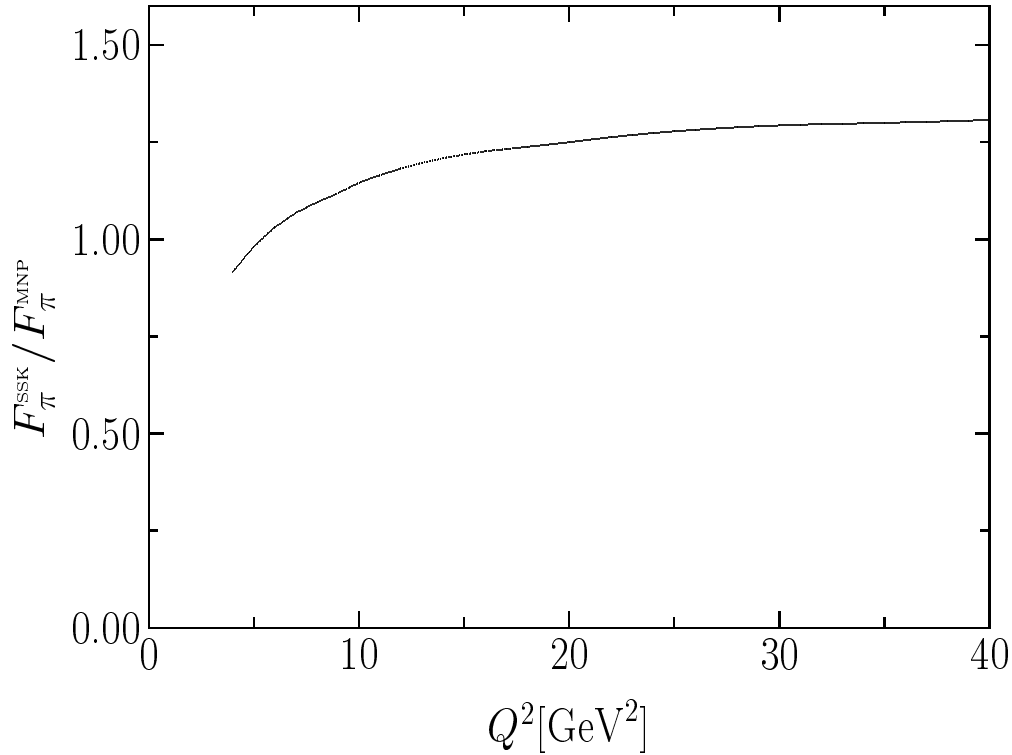,height=10cm,width=14cm,silent=}}
\vspace{1cm}
\caption[fig:pionratio]
        {\tenrm The ratio
         $
          R_{\pi}\left(Q^{2}\right)
           \equiv
          F_{\pi}^{\rm SSK}\left(Q^{2}\right)/
          F_{\pi}^{\rm MNP}\left(Q^{2}\right)
         $
         versus the momentum transfer $Q^{2}$.
         $F_{\pi}^{\rm SSK}$ is the pion form factor given in
         Eq.~(\ref{eq:NLOpiff}), $F_{\pi}^{\rm MNP}$ the expression
         derived in \cite{MNP98} (cf.~Table \ref{tab:piffval}), using
         the asymptotic pion distribution amplitude. The decrease of
         this ratio below $Q^{2}\simeq 5$~GeV${}^{2}$ signals the
         breakdown of perturbation theory in the calculation of
         \cite{MNP98} owing to the Landau singularity in the
         conventional $\alpha _{\rm s}$ representation they use.
\label{fig:ratioSSKMNP}}
\end{figure}

The illustration of the enhanced form-factor behavior (always assuming
the asymptotic form of the pion distribution amplitude) is given in
Fig.~\ref{fig:ratioSSKMNP} in terms of the ratio between
$F_{\pi}^{\rm SSK}\left(Q^{2}\right)$, calculated in this work, and
$F_{\pi}^{\rm MNP}\left(Q^{2}\right)$, obtained by Meli\v c
{\it et al.} in \cite{MNP98}.
One sees from that figure that at $Q^{2}$ values up to about
5~GeV${}^{2}$, this ratio is less than unity, clearly exhibiting the
singular IR-behavior of the conventional $\alpha _{\rm s}(Q^{2})$
representation, employed by these authors.
Contrary to that, above approximately 10 GeV${}^{2}$, this ratio
scales with $Q^{2}$ at a fixed value of about $1.25$.
Hence, restoring analyticity of the effective QCD coupling (by a power
correction term), removes the artificial raise of the form factor, owing
to the rapid increase of the perturbative coupling at low momentum, and
stabilizes its low-$Q^{2}$ behavior, providing enhancement only in that
momentum region which is controlled by self-consistent perturbation
theory.

%%%%%%%%%%%%%%%%%%%%%%%%%%%%%%%%%%%%%%%%%%%%%%%%%%%%%%%%%%%%%%%%%%%%%
%                           F I G U R E  8                          %
%                      \label{fig:ratioSSKMNP}                      %
%%%%%%%%%%%%%%%%%%%%%%%%%%%%%%%%%%%%%%%%%%%%%%%%%%%%%%%%%%%%%%%%%%%%%
%
%%%%%%%%%%%%%%%%%%%%%%%%%%%%%%%%%%%%%%%%%%%%%%%%%%%%%%%%%%%%%%%%%%%%%

Let us turn again to the calculation of $F_{\pi\gamma}(Q^{2})$.
Figure \ref{fig:pigaepj} shows our theoretical predictions for this
form factor using the same set of scheme parameters
$C_{1}, C_{2}, C_{3}$, given in Table \ref{tab:setsC_i}.
The dashed line includes a quark mass term and employs commensurate
scale setting for the renormalization point.
The solid line shows the prediction for $m_{\rm q}=0$ and a
non-commensurate renormalization scale, with
$C_{4}=C_{2}=\exp{(-1/2)}$.
This latter curve reproduces the recent high-precision CLEO
\cite{CLEO98} and also the earlier CELLO \cite{CELLO91} data with almost
the same numerical accuracy as the dipole interpolation formula.
However, we regard the lower curve as being more realistic because
a physical renormalization scale has been used (provided our choice
of $C_{4}=C_{2}\exp{(-1/2)}$ is approximately correct).
Remarkably, the predicted magnitude of $Q^{2}F_{\pi\gamma}$, being
somewhat below the data, allows some broadening of the pion distribution
amplitude, as recently found in instanton-based approaches
\cite{Dor96,PPRWG98} or using non-local condensates \cite{MR86,BM95}.

\begin{figure}
\tighten
\centerline{\epsfig{file=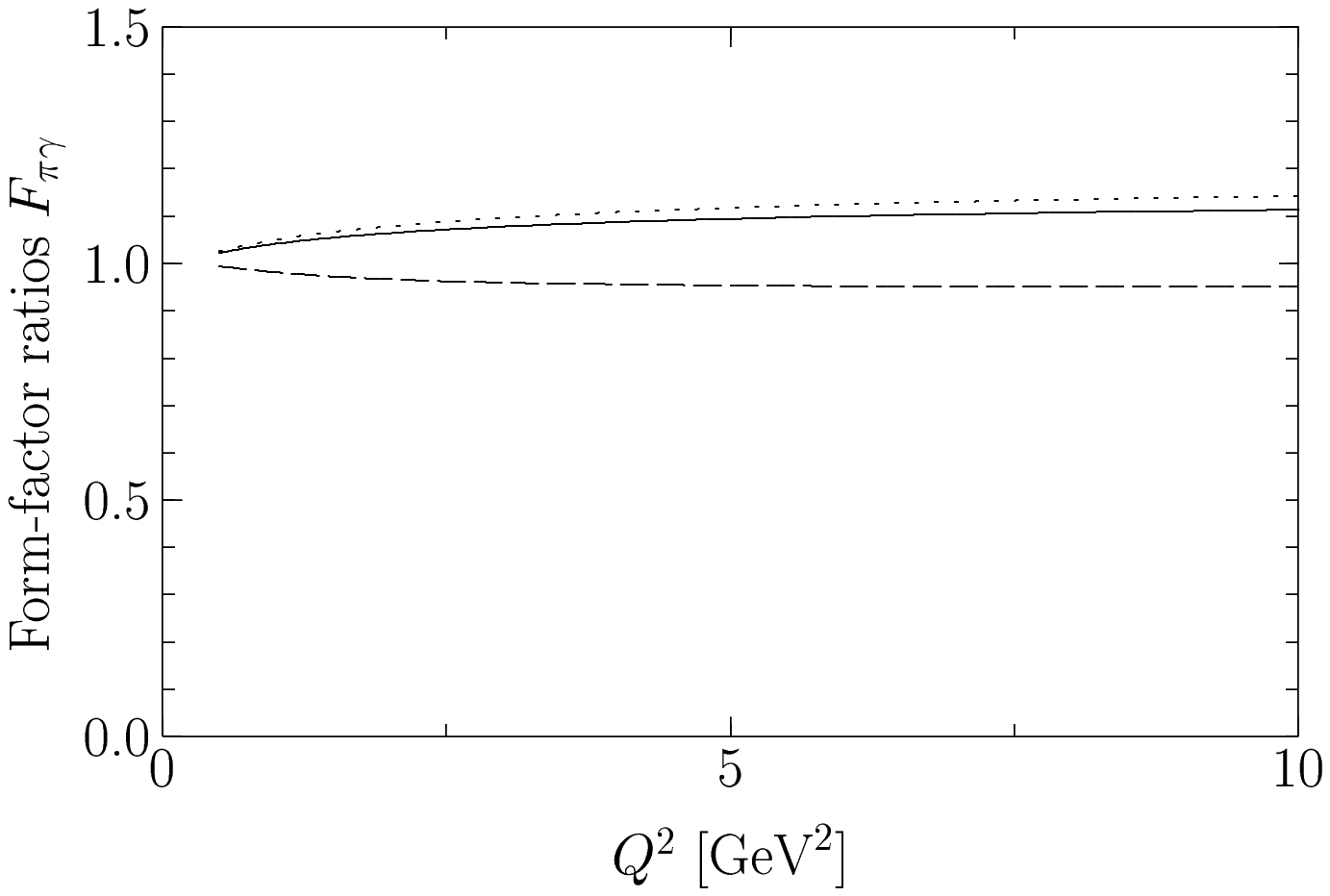,height=8cm,width=13cm,silent=}}
\vspace{1cm}
\caption[fig:div_pigaeffects]
        {\tenrm Effects of commensurate scale fixing and a non-zero
                quark mass in the BHL ansatz for the pion wave
                function in the pion-photon transition form factor.
                The solid line shows the ratio
$
 {\left(F_{\pi\gamma}^{m_{\rm q}=0}/
        F_{\pi\gamma}^{m_{\rm q}\neq 0}\right)}\big|_{\rm comm}
$
                for a commensurate scale fixing and the dotted line
                the same ratio for a non-commensurate scale fixing:
$
 {\left(F_{\pi\gamma}^{m_{\rm q}=0}/
        F_{\pi\gamma}^{m_{\rm q}\neq 0}\right)}\big|_{\rm non-comm}
$
                with
                $C_{4}=C_{2}=\exp{(-1/2)}$.
                The dashed line effects the difference between using a
                commensurate scale fixing and a conventional one
                with $C_{4}=C_{2}=\exp{(-1/2)}$ for the ratio
$
 {\left(F_{\pi\gamma}^{\rm comm}/
        F_{\pi\gamma}^{\rm non-comm}\right)}\big|_{m_{\rm q}\neq 0}
$
                with a non-vanishing quark mass $m_{\rm q}=0.33$~GeV.
\label{fig:rapiga}}
\end{figure}

The sensitivity of the pion-photon transition form factor to the quark
mass and the commensurate scale setting is discussed in
Fig.~\ref{fig:rapiga}.

%%%%%%%%%%%%%%%%%%%%%%%%%%%%%%%%%%%%%%%%%%%%%%%%%%%%%%%%%%%%%%%%%%%%%
%                           F I G U R E  9                          %
%                         \label{fig:rapiga}                        %
%%%%%%%%%%%%%%%%%%%%%%%%%%%%%%%%%%%%%%%%%%%%%%%%%%%%%%%%%%%%%%%%%%%%%
%
%%%%%%%%%%%%%%%%%%%%%%%%%%%%%%%%%%%%%%%%%%%%%%%%%%%%%%%%%%%%%%%%%%%%%

%%%%%%%%%%%%%%%%%%%%%%%%%%%%%%%%%%%%%%%%%%%%%%%%%%%%%%%%%%%%%%%%%%%%%
%                           F I G U R E  10                         %
%                       \label{fig:ratioalpha}                      %
%%%%%%%%%%%%%%%%%%%%%%%%%%%%%%%%%%%%%%%%%%%%%%%%%%%%%%%%%%%%%%%%%%%%%
\begin{figure}
\tighten
\centerline{\epsfig{file=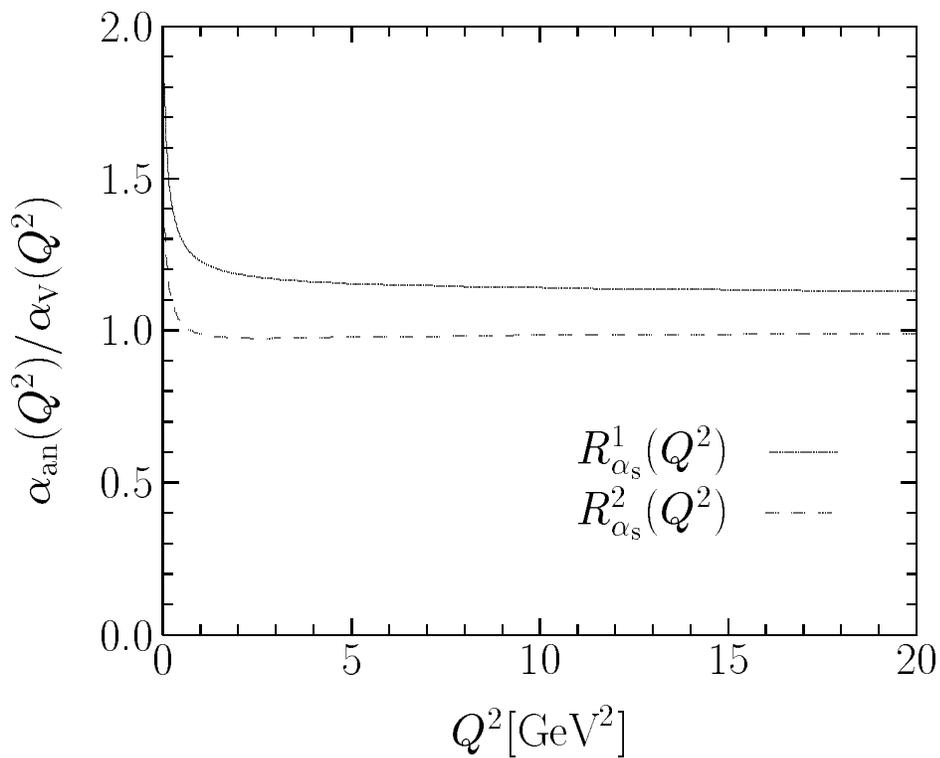,height=10cm,width=14cm,silent=}}
\vspace{1cm}
\caption[fig:ratio_an_v]
        {\tenrm The ratio
         $
           R_{\alpha _{\rm s}^{1}}\left(Q^{2}\right)
          \equiv
           \alpha _{\rm an}^{\rm s}\left(Q^{2}\right)/
           \alpha _{\rm V}^{\rm s}\left(Q^{2}\right)
         $
         versus the momentum transfer $Q^{2}$.
         $\alpha _{\rm an}^{\rm s}$ is the analytic running coupling
         in one-loop approximation (see,
         Eq.~(\ref{eq:oneloopalpha_an})).
         The effective charge $\alpha _{\rm V}^{\rm s}$, used by Brodsky
         et al.~in \cite{BJPR98}, is defined in
         Eq.~(\ref{eq:alpha_V}).
         For more explanations and the definition of
         $R_{\alpha _{\rm s}^{2}}\left(Q^{2}\right)$,
         see in the text.
\label{fig:ratioalpha}}
\end{figure}
%
%%%%%%%%%%%%%%%%%%%%%%%%%%%%%%%%%%%%%%%%%%%%%%%%%%%%%%%%%%%%%%%%%%%%%

%%%%%%%%%%%%%%%%%%%%%%%%%%%%%%%%%%%%%%%%%%%%%%%%%%%%%%%%%%%%%%%%%%%%%
%                             T A B L E  4                          %
%                         \label{tab:piffmom}                       %
%%%%%%%%%%%%%%%%%%%%%%%%%%%%%%%%%%%%%%%%%%%%%%%%%%%%%%%%%%%%%%%%%%%%%
\begin{table}
\caption[tab:pionffvalues]
        {Values of the scaled space-like pion form factor, calculated in
         our IRF scheme at different momentum transfers $Q^{2}$.
         $Q^{2}F_{\pi}^{\rm LO}\left(Q^{2}\right)$ is the LO result
         given by Eq.~(\ref{eq:pifofafin}) and represented by the
         dashed-dotted line in Fig.~\ref{fig:piffepj}.
         $Q^{2}F_{\pi}^{\rm NLO}\left(Q^{2}\right)$ (dotted line in
         Fig.~\ref{fig:piffepj}) is the expression displayed in
         Eq.~(\ref{eq:NLOpiff}) and comprises the LO and NLO
         contributions to the hard-scattering part (for more details,
         see Sect.~\ref{sec:piffNLO}). These results were obtained with
         a non-factorizing BHL-type ansatz for the pion wave function
         (i.e., with an effective (constituent-like) quark mass
         $m_{\rm q}=0.33$~GeV in the pion wave function), and employing
         BLM commensurate (renormalization) scale setting.
         The last two columns show the results for the pion-photon
         transition. $Q^{2}F_{\pi\gamma}\left(Q^{2}\right)$ stands
         for the expression (\ref{eq:ffpigamma}) and commensurate scale
         setting, whereas
         $Q^{2}F_{\pi\gamma}^{m_{\rm q}=0}\left(Q^{2}\right)$
         shows the results without the inclusion of a quark mass and
         with a non-commensurate renormalization scale (cf.~dashed and
         solid lines in Fig.~\ref{fig:pigaepj}, respectively). The
         asymptotic pion distribution amplitude is always assumed.
\label{tab:piffmom}}
\begin{tabular}{cccccc}
 $Q^{2}$~[GeV${}^{2}$] & $Q^{2}F_{\pi}^{\rm LO}\left(Q^{2}\right)$
                       & $Q^{2}F_{\pi}^{\rm NLO}\left(Q^{2}\right)$
                       & $Q^{2}F_{\pi\gamma}^{}\left(Q^{2}\right)$
                & $Q^{2}F_{\pi\gamma}^{m_{\rm q}=0}\left(Q^{2}\right)$
                       \\
\hline
2   & 0.1121 & 0.1831 & 0.1180 & 0.1370 \\
4   & 0.1282 & 0.1907 & 0.1317 & 0.1576 \\
6   & 0.1340 & 0.1904 & 0.1375 & 0.1668 \\
8   & 0.1364 & 0.1882 & 0.1407 & 0.1721 \\
10  & 0.1373 & 0.1856 & 0.1428 & 0.1755 \\
15  & 0.1368 & 0.1790 & 0.1458 & 0.1803 \\
20  & 0.1351 & 0.1731 & 0.1474 & 0.1828 \\
30  & 0.1312 & 0.1639 & 0.1491 & 0.1854 \\
40  & 0.1275 & 0.1568 & 0.1500 & 0.1867 \\
\end{tabular}
\end{table}
%T A B L E  4
%
%%%%%%%%%%%%%%%%%%%%%%%%%%%%%%%%%%%%%%%%%%%%%%%%%%%%%%%%%%%%%%%%%%%%%

Our prediction is consistent with the result obtained by Brodsky
{\it et al.} in \cite{BJPR98}, who also use commensurate scale setting
and include in addition the LO QCD radiative correction to
$F_{\pi\gamma}$ with a running coupling ``frozen'' at low momenta by
virtue of an effective gluon mass.\footnote{The connection between the
modified convolution scheme, which explicitly retains transverse
degrees of freedom, and the use of an effective gluon mass to simulate
the effect of the Sudakov suppression factor, was discussed in
\cite{Ste99}.}
The close resemblance between the two approaches becomes apparent by
comparing the corresponding running couplings against the momentum
transfer.
In Figure \ref{fig:ratioalpha} we show the ratio (solid line)
\begin{equation}
  R_{\alpha _{\rm s}^{1}}\left(Q^{2}\right)
\equiv
  \frac{\alpha _{\rm an}^{\rm s}\left(Q^{2}\right)}
  {\alpha _{\rm V}^{\rm s}\left(Q^{2}\right)}
\label{eq:ratalph}
\end{equation}
%Eq (50) Ratio of analytic coupling and coupling in the V-scheme
with $\alpha _{\rm an}^{\rm s}\left(Q^{2}\right)$ given by
Eq.~(\ref{eq:oneloopalpha_an}) and the coupling (effective charge) in
the so-called V scheme, defined by
\begin{equation}
  \alpha _{\rm V}^{\rm s}\left(Q^{2}\right)
=
  \frac{4\pi}{\beta _{0}}\,\frac{1}{
  \ln \left(\frac{Q^{2} + 4m_{\rm g}^{2}}{\Lambda _{\rm V}^{2}}\right)}
\; ,
\label{eq:alpha_V}
\end{equation}
%Eq (51) Effective coupling (charge) in V scheme
where
$\Lambda _{\rm V}=0.16$ \cite{BJPR98} and
$m_{\rm g}^{2}=0.19$~GeV${}^{2}$.
The dashed line ($R_{\alpha _{\rm s}^{2}}\left(Q^{2}\right)$) represents
this ratio with $\Lambda _{\rm V}$ set equal to
$\Lambda = 0.242$~GeV in Eq.~(\ref{eq:alpha_V}).
Though, strictly speaking, it is inconsistent to equalize
scheme-dependent parameters, the message of this figure is that
the two parameterizations are very close to each other, albeit the
analytic coupling has a larger normalization at low $Q^{2}$.

Closing our discussion of the photon to pion transition, let us mention
that other authors \cite{RR96,MR97,Kho97} obtain similarly good numerical
agreement of $Q^{2}F_{\pi\gamma ^{*}}\left(Q^{2}\right)$ with the
experimental data, following different premises based on QCD sum rules.

Finally to facilitate a more detailed comparison of our results with
other approaches and experimental data, we compile in
Table \ref{tab:piffmom} the obtained values of the (scaled) pion
electromagnetic (LO and NLO) and photon to pion (LO) form factors at
different momentum transfers $Q^{2}$.
These form factors are calculated with a non-factorizing BHL-type
ansatz for the pion wave function (hence including an effective
quark mass in the Gaussian distribution for the intrinsic transverse
momentum) and using a BLM commensurate fixing of the renormalization
scale.
In the case of the pion-photon transition form factor, we also show the
result setting the constituent quark mass equal to zero and employing a
non-commensurate renormalization scale -- in analogy to our previous
analysis in \cite{SSK98}.
In all cases, the asymptotic form of the pion distribution amplitude is
assumed.

\section{S\lowercase{ummary and conclusions}}
\label{sec:sum}

Let us summarize the hallmarks of the presented methodology.
We have developed in detail a theoretical framework which
self-consistently incorporates effects resulting from a modification
of the strong running coupling by a non-perturbative minimum power
correction \cite{SS97} which provides IR universality.
Though a deep physical understanding of such contributions is still
lacking, we have given, as a matter of practice, quantitative evidence
that using such an analytic running coupling it is possible to get an
IR-enhanced hard contribution to the electromagnetic form factor
$F_{\pi}(Q^{2})$ by employing only asymptotic (like) forms of the pion
distribution amplitude, hence without recourse to end-point concentrated
distribution amplitudes.

The presented IR-finite factorization and renormalization scheme makes
it possible to take into account transverse degrees of freedom both in
the pion wave function \cite{JK93} as well as in the form of Sudakov
damping factors \cite{LS92}, without entailing suppression of the
(pion) form-factor magnitude resulting from severe IR regularization.
In addition, use of this modified form of $\alpha _{\rm s}(Q^{2})$
renders the theoretical predictions insensitive to its variation with
$Q^{2}$ at small momentum values, thus remarkably improving their
scaling behavior, in accordance to the nominal scaling of the leading
twist contribution.
Similarly, the saturation behavior of the pion form factor (versus
the impact separation) is significantly improved and the scaled hard
form factor reaches much faster a plateau, accumulating its magnitude
in the region of small transverse distances where use of perturbation
theory is legal.
An appropriate choice of the factorization (scheme) scales and the
strict separation between gluonic contributions from fixed-order and
re-summed perturbation theory helps avoid double counting of
higher-order corrections, enforcing this way the self-consistency of
the whole perturbative treatment in a wide range of momentum transfer.
Moreover, adopting the BLM commensurate procedure in order to choose
an optimized renormalization scale, and thus minimize the
renormalization scheme dependence, we have calculated the pion form
factor including the NLO radiative correction to the hard-scattering
amplitude.
In contrast to other approaches, we employ a BHL-type of ansatz for the
distribution of the intrinsic transverse momentum in the pion wave
function which includes a mass term.
This term, resulting from the nonperturbative QCD vacuum structure,
ensures suppression of
$\Psi _{\pi}^{\rm soft}\left(x,{\kv}\right)$
for ${\kv}=0$ and $k_{3}\to -\infty$,
and yields to a stronger fall-off of the soft, non-factorizing
contribution to the form factor at momentum-transfer values around
20~GeV${}^{2}$ and beyond.
Hence, the leading-twist predictions of QCD are remarkably confirmed
at still higher $Q^{2}$, whereas at lower momentum values Feynman-type
contributions dominate.
In this region other higher-twist contributions may also be important.

The same procedure applied, without any scheme parameter re-tuning,
to $F_{\pi ^{0}\gamma ^{*}\gamma}$ yields a prediction which is
consistent with, though somehow below, the experimental data of the
CLEO and CELLO collaborations, and allows therefore for a mild
broadening of the (true) pion wave function, as indicated by
instanton-based approaches.

We believe that the insight gained through our analysis gives a strong
argument that a power correction in the running coupling of QCD, as
proposed by Shirkov and Solovtsov, has important consequences and
provides a convenient tool to improve theoretical predictions based on
perturbation theory.

\bigskip

\ac
We wish to thank Alexander Bakulev, Anatoly Efremov, Rainer Jakob,
Alekos Karanikas, and Anatoly Radyushkin for discussions and Stan
Brodsky for useful communications.
The work of H.-Ch.K. was supported by the Research Institute for
Basic Sciences, Pusan National University under Grant RIBS-PNU-99-203.
\newpage        %finishes textpart
\newpage        %finishes acknowledgment

\newpage        %finishes references
\end{document}